\documentclass[twocolumn,aps,prl]{revtex4}
\usepackage{amsmath}
\usepackage{graphicx}
\usepackage[british]{babel}

\graphicspath{{./graphics/}}
\newcommand{\MSlabel}[1]{\tag{A\theequation}\label{#1}\stepcounter{equation}}

\begin{document}


\title{Theory of adhesion: role of surface roughness}
\author{B.N.J. Persson$^1$ and M. Scaraggi$^{1,2}$}
\affiliation{$^1$PGI, FZ-J\"ulich, 52425 J\"ulich, Germany, EU}
\affiliation{$^2$DII, Università del Salento, 73100 Monteroni-Lecce, Italy, EU}

\begin{abstract}
We discuss how surface roughness influence the adhesion between elastic
solids. We introduce a Tabor number which depends on the length scale or
magnification, and which gives information about the nature of the adhesion
at different length scales. We consider two limiting cases relevant for (a)
elastically hard solids with weak adhesive interaction (DMT-limit) and (b)
elastically soft solids or strong adhesive interaction (JKR-limit). For the
former cases we study the nature of the adhesion using different adhesive
force laws ($F\sim u^{-n}$, $n=1.5-4$, where $u$ is the wall-wall
separation). In general, adhesion may switch from DMT-like at short length
scales to JKR-like at large (macroscopic) length scale. 
We compare the theory
predictions to the results of exact numerical simulations and find good
agreement between theory and the simulation results.
\end{abstract}

\maketitle



\textbf{1 Introduction}

Surface roughness has a huge influence on the adhesion and friction between
macroscopic solid objects\cite{Bowden,Johnson,Isra,BookP,R1,P3}. Most
interaction force fields are short ranged and becomes unimportant when the
separation between solid surfaces exceed a few atomic distances, i.e., at
separations of order nm. This is trivially true for chemical bonds (covalent
or metallic bonds) but holds also for the more long-ranged Van der Waals
interaction. One important exception is charged bodies. For uncharged
solids, if the surface roughness amplitude is much larger than the decay
length of the wall-wall interaction potential and if the solids are
elastically stiff enough, no macroscopic adhesion will prevail, as is the
case in most practical cases. Only for very smooth surfaces, or elastically
very soft solids (which can deform and make almost perfect contact at the
contacting interface without storing up a large elastic energy) adhesion will
be observed for macroscopic solids\cite{Fuller}.

In this paper we will discuss how surface roughness influence adhesion
between macroscopic solids. We consider two limiting cases, which are valid
for elastically hard and weakly interaction solids (Deryagin, Muller, and
Toporov, DMT-limit)\cite{DMTP} and for elastically soft or strongly
interacting solids (Johnson, Kendall, and Roberts, JKR-limit)\cite{JKRP}.
This problem has been studied before but usually using the
Greenwood-Williamson\cite{GW,JG} type of asperity models (see, e.g., \cite%
{Fuller,Maugis96}), whereas our treatment is based on the Persson contact
mechanics model. The latter model is (approximately) valid even close to
complete contact (which often prevail when adhesion is important)\cite{Mus1,Mus3}. 
Asperity models can only be used as long as the contact area is small
compared to the nominal contact area, and even in this limit these models
have severe problems for surfaces with roughness on many length scales\cite%
{Carb1,Comb1,Problem}.

Recently several numerical studies of adhesion between randomly rough
surfaces have been published. Pastewka and Robbins\cite{mark} study the
adhesion between rough surfaces and present a criterion for macroscopic
adhesion. They emphasize the role of the range of the adhesive interaction,
which we also find is important in the DMT limit and when the surface
roughness amplitude is small (see below). Medina and Dini\cite{dini} studied
the adhesion between an elastic sphere with smooth surface and a rigid
randomly rough substrate surface. They observed strong contact hysteresis in
the JKR-limit (relative smooth surfaces) and very small contact hysteresis
in the DMT-limit which prevails for small roughness. Analytical theories of
contact mechanics have been compared to numerically exact calculations for
two-dimensional (2D) randomly rough surfaces in Ref. \cite{MP} and for 1D
surface roughness in Ref. \cite{CarboneS}. Experimental adhesion data for
rough surfaces have been compared to analytical theory predictions in Ref. 
\cite{PerssonPRL}, \cite{Krick} and \cite{Krick1}.

Many practical or natural adhesive systems involve effects which usually
are not considered in model studies of adhesion, and which we will not
address in this paper. In particular, biological applications typically
involve complex structured surfaces (e.g., hierarchical fiber-and-plate
structures) with anisotropic elastic properties, which are elastically soft
on all relevant length scales\cite{Aut,Heepe,PerG1,PerG2}. Instead of
directly relying on molecular bonding over atomic dimension, many biological
systems adhere mainly via capillary bridges\cite{cellulose,skin,frog}. We
will also not discuss either the adhesion between charged objects, which must be
treated by special methods which takes into account the long-range nature of
the Coulomb interaction\cite{coulomb3,coulomb4,coulomb1,coulomb2}.

In this paper we first briefly review (Sec. 2) two limiting models of
adhesion for smooth surfaces. In Sec. 3 we show how the same limiting cases
can be studied analytically for randomly rough surfaces using the Persson
contact mechanics model. Numerical results obtained using the analytical
theory are presented in Sec. 4, and compared to exact numerical results in
Sec. 5. Sec. 6 contains a discussion and Sec. 7 the summary and conclusion. 
\begin{figure}[tbp]
\includegraphics[width=0.45\textwidth]{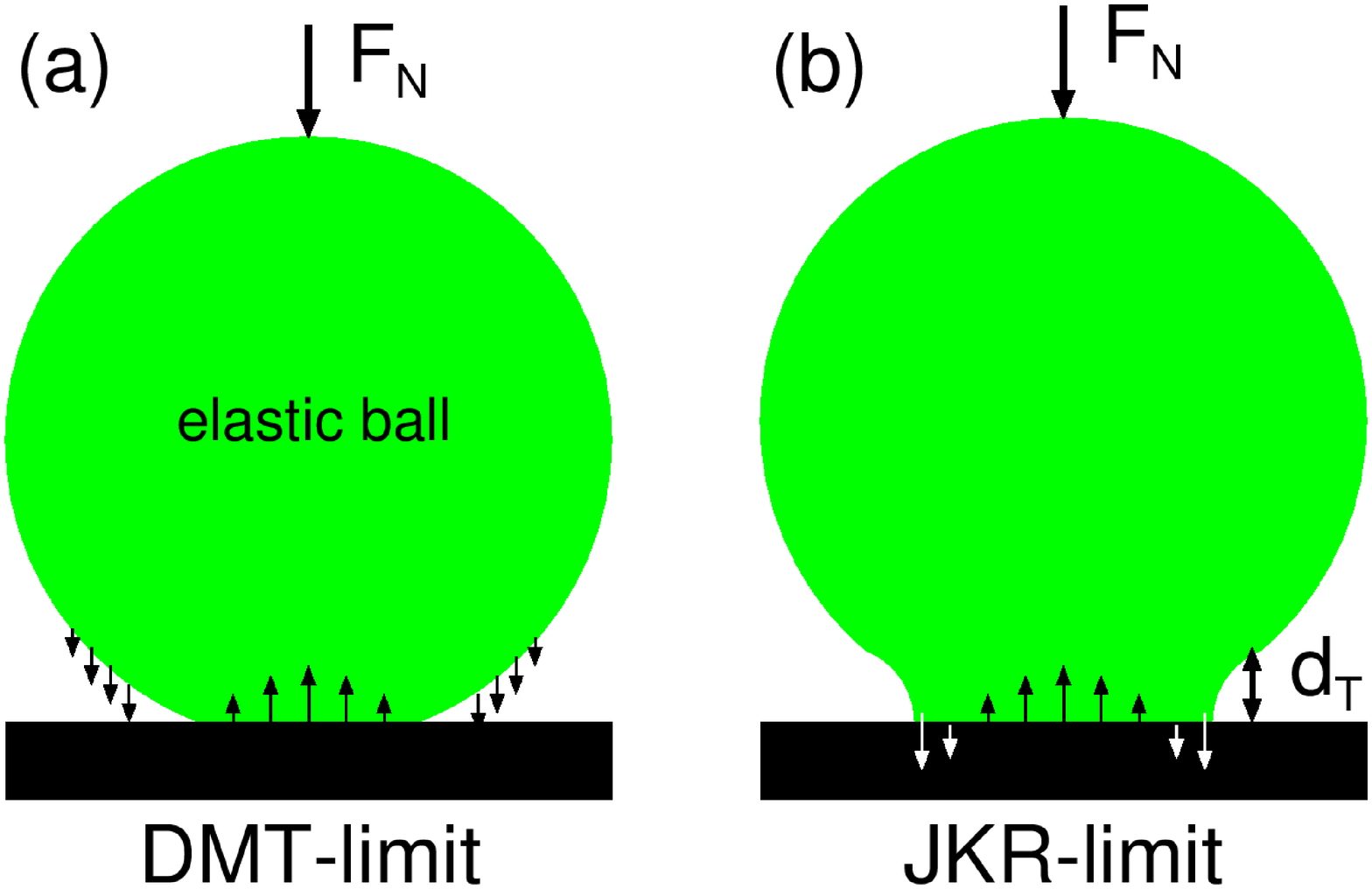}
\caption{(a) In the DMT theory the elastic deformation field is calculated
with the adhesion included only as an additional load $F_{\mathrm{ad}}$
acting on the sphere. Thus the contact area is determined by Hertz theory
with the external load $F_{\mathrm{N}}+F_{\mathrm{ad}}$. The adhesional load 
$F_{\mathrm{ad}}$ is obtained by integrating the adhesional stress over the
ball non-contact area. (b) In the JKR theory the adhesion force is assumed
to have infinitesimal spatial extent, and is included only in the contact
area as an interfacial binding energy $E_{\mathrm{ad}}=\Delta \protect\gamma %
A$. The shape of the elastic body is obtained by minimizing the total energy 
$-E_{\mathrm{ad}}+U_{\mathrm{el}}$, where $U_{\mathrm{el}}$ is the elastic
deformation energy. }
\label{ballshape.eps}
\end{figure}

\vskip 0.3cm \textbf{2 Adhesion of ball on flat (review)}

Analytical studies of adhesion have been presented for smooth surfaces for
bodies of simple geometrical shape, the most important case being the
contact between spherical bodies. For a sphere in contact with a flat
surface two limiting cases are of particular importance, usually referred to
as the DMT theory\cite{DMTP} and the JKR theory\cite{JKRP}, see Fig. \ref%
{ballshape.eps}. Analytical results for intermediate-range adhesion was
presented by Maugis\cite{Dug,Bart} and the ball-flat adhesion problem has
also been studied in detail using numerical methods\cite{Der,Gree}. A
particular detailed numerical study was recently published by M\"user who
also included negative work of adhesion (repulsive wall-wall interaction)%
\cite{Mart}.

Consider an elastic ball (e.g., a rubber ball) with the radius $R$, Young's
elastic modulus $E$ (and Poisson ratio $\nu$), in adhesive contact with a
flat rigid substrate. Let $\Delta \gamma = \gamma_1+\gamma_2-\gamma_{12}$ be
the work of adhesion and let $d_{\mathrm{c}}$ be the spatial extend of the
wall-wall interaction potential (typically of order atomic distance). The
DMT theory is valid when adhesive stress $\sigma_{\mathrm{ad}} \approx
\Delta \gamma /d_{\mathrm{c}}$ is much smaller than the stress in the
contact region, which is of order 
\begin{equation*}
\sigma_{\mathrm{c}} \approx \left ({\frac{\Delta \gamma E^2 }{R}}\right
)^{1/3}.
\end{equation*}
In the opposite limit the JKR theory is valid. In the DMT theory the elastic
deformation field is calculated with the adhesion included only as an
additional load $F_{\mathrm{ad}}$ acting on the sphere. Thus the contact
area is determined by Hertz theory with the external load $F_0=F_{\mathrm{N}%
}+F_{\mathrm{ad}}$, where $F_{\mathrm{N}}$ is the actual load on the ball
(see Fig. \ref{ballshape.eps}). The adhesion load $F_{\mathrm{ad}}$ is
obtained by integrating the adhesion stress over the ball non-contact area.

The JKR theory neglects the extend of the interaction potential and assumes
interaction between the solids only in the contact area. The deformation
field in the JKR theory is obtained by minimizing the total energy given by
the sum of the (repulsive) elastic deformation energy and the (attractive)
binding energy $E_{\mathrm{ad}} = \Delta \gamma A$, where $A$ is the contact
area. In this theory the contribution to binding energy from the non-contact
region is neglected.

Since $\sigma_{\mathrm{ad}} \approx \Delta \gamma / d_{\mathrm{c}}$ we can
define the Tabor number: 
\begin{equation*}
\mu_{\mathrm{T}} = {\frac{\sigma_{\mathrm{ad}} }{\sigma_{\mathrm{c}}}} =
\left ( {\frac{R\Delta \gamma^2 }{E_{\mathrm{r}}^2 d_{\mathrm{c}}^3 }}\right
)^{1/3} = {\frac{d_{\mathrm{T}} }{d_{\mathrm{c}}}},
\end{equation*}
where 
\begin{equation*}
d_{\mathrm{T}} = \left ( {\frac{R\Delta \gamma^2 }{E_{\mathrm{r}}^2 }}\right
)^{1/3},
\end{equation*}
where $E_{\mathrm{r}}=E/(1-\nu ^{2})$ is effective elastic modulus. The DMT
and JKR limits correspond to $\mu_{\mathrm{T}} << 1$ and $\mu_{\mathrm{T}}
>>1$, or, equivalently, $d_{\mathrm{T}} << d_{\mathrm{c}}$ and $d_{\mathrm{T}%
} >> d_{\mathrm{c}}$, respectively. In the JKR-limit the Tabor length $d_{%
\mathrm{T}}$ can be considered as the height of the neck which is formed at
the contact line (see Fig. \ref{ballshape.eps}(b)). This neck height must be
much larger than the length $d_{\mathrm{c}}$, which characterizes the spatial
extend of the wall-wall interaction, in order for the JKR-limit to prevail.

At vanishing external load, $F_{\mathrm{N}}=0$, the JKR theory predicts the
contact area: 
\begin{equation*}
A_{\mathrm{JKR}}=\pi \left( {\frac{9\pi R^{2}\Delta \gamma }{2E_{\mathrm{r}}}%
}\right) ^{2/3}.
\end{equation*}%
This contact area is a factor $3^{2/3}\approx 2.1$ larger than obtained from
the DMT theory. In the JKR theory the force necessary to remove the ball
from the flat (the pull-off force) is given by 
\begin{equation}
F_{\mathrm{c}}={\frac{3\pi }{2}}\Delta \gamma R  \label{1}
\end{equation}%
which is a factor of $3/4$ times smaller than predicted by the DMT theory.
Also the pull-off process differs: in the JKR theory an elastic instability
occurs where the contact area abruptly decreases, while in the DMT theory the
contact area decreases continuously, until the ball just touches the substrate
in a single point, at which point the pull-force is maximal.

For the sphere-flat case the pull-off force in the DMT-limit is \textit{%
independent} of the range of the wall-wall interaction potential. However,
this is not the case for other geometries where in fact the contact
mechanics depends remarkably sensitively on the interaction range. As a
result the interaction between rough surfaces in the DMT-limit will depend
on the force law as we will demonstrate below for power law interaction $p_{%
\mathrm{ad}} \sim u^{-n}$.

In an exact treatment, as a function of the external load $F_{\mathrm{N}}$,
the total energy $E_{\mathrm{tot}}=-E_{\mathrm{ad}}+U_{\mathrm{el}}$ must
have a minimum at $F_{\mathrm{N}} = 0$. This is the case in the JKR theory
but in general not for the DMT theory. However, the DMT theory is only valid
for very stiff solids and in this limiting case the total energy minimum
condition is almost satisfied. Nevertheless, one cannot expect $dE_{\mathrm{%
tot}} /dF_{\mathrm{N}} (F_{\mathrm{N}}=0)=0$ to be exactly obeyed in any
(approximate) theory which does not focus on minimizing the total energy.

The results above assume perfectly smooth surfaces. The JKR (and DMT) theory results
can, however, be applied also to surfaces with roughness assuming that the
wavelength $\lambda$ of the most longest (relevant) surface roughness
component is smaller than the diameter of the contact region. In that case
one only needs to replace the work of adhesion $\Delta \gamma$ for flat
surfaces with an effective work of adhesion $\gamma_{\mathrm{eff}}$ obtained
for the rough surfaces. We will now describe how one may calculate $\gamma_{%
\mathrm{eff}}$.

\vskip 0.3cm \textbf{3 Theory: basic equations}

We now show how surface roughness can be taken into account in adhesive
contact mechanics. We consider two limiting cases similar to the JKR and DMT
theories for adhesion of a ball on a flat. The theory presented below is not
based on the standard Greenwood-Williamson\cite{GW,JG} picture involving
contact between asperities, but on the Persson contact mechanics theory.

\vskip 0.3cm \textbf{3.1 JKR-limit}

In the JKR-limit the spatial extend of the wall-wall interaction potential
is neglected so the interaction is fully characterized by the work of
adhesion $\Delta \gamma$.

In order for two elastic solids with rough surfaces to make adhesive contact
it is necessary to deform the surfaces elastically, otherwise they would
only make contact in three points and the adhesion would vanish, at least if
the spatial extend of the adhesion force is neglected. Deforming the
surfaces to increase the contact area $A$ results in some interfacial bonding 
$-\Delta \gamma A$ (where $\Delta \gamma = \gamma_1+\gamma_2-\gamma_{12}$ is
the change in the interfacial energy per unit area upon contact), but it
costs elastic deformation energy $U_{\mathrm{el}}$, which will reduce the
effective binding. That is, during the removal of the block from the
substrate the elastic compression energy stored at the interface is given
back and helps to break the adhesive bonds in the area of real contact. Most
macroscopic solids do not adhere with any measurable force, which implies
that the total interfacial energy $-\Delta \gamma A + U_{\mathrm{el}}$
vanishes, or nearly vanishes, in most cases.

The contact mechanics theory of Persson\cite%
{P1,P2,P3,P4,YP,Carlos,Layer,Carbone} can be used to calculate
(approximately) the stress distribution at the interface, the area of real
contact and the interfacial separation between the solid walls\cite{P1,P4}.
In this theory the interface is studied at different magnifications $\zeta
=L/\lambda $, where $L$ is the linear size of the system and $\lambda $ the
resolution. We define the wavevectors $q=2\pi /\lambda $ and $q_{0}=2\pi /L$
so that $\zeta =q/q_{0}$. The theory focuses on the probability distribution $%
P(\sigma ,\zeta )$ of stresses $\sigma $ acting at the interface when the
system is studied at the magnification $\zeta $. In Ref. \cite{P1} an
approximate diffusion equation of motion was derived for $P(\sigma ,\zeta )$%
. To solve this equation one needs boundary conditions. If we assume that,
when studying the system at the lowest magnification $\zeta =1$ (where no
surface roughness can be observed, i.e., the surfaces appear perfectly
smooth), the stress at the interface is constant and equal to $p_{\mathrm{N}%
}=F_{\mathrm{N}}/A_{0}$, where $F_{\mathrm{N}}$ is the load and $A_{0}$ the
nominal contact area, then $P(\sigma ,1)=\delta (\sigma -p_{\mathrm{N}})$.
In addition to this \textquotedblleft initial condition\textquotedblright\
we need two boundary conditions along the $\sigma $-axis. Since there can be
no infinitely large stress at the interface we require $P(\sigma ,\zeta
)\rightarrow 0$ as $\sigma \rightarrow \infty $. For adhesive contact, which
interests us here, tensile stress occurs at the interface close to the
boundary lines of the contact regions. In this case we have the boundary
condition $P(-\sigma _{\mathrm{a}},\zeta )=0$, where $\sigma _{\mathrm{a}}>0$
is the largest (locally averaged at magnification $\zeta $) tensile stress
possible. Hence, the detachment stress $\sigma _{\mathrm{a}}(\zeta )$
depends on the magnification and can be related to the effective interfacial
energy (per unit area) $\gamma _{\mathrm{eff}}(\zeta )$ using the theory of
cracks\cite{P3}. The effective interfacial binding energy%
\begin{equation*}
\gamma _{\mathrm{eff}}(\zeta )A(\zeta )=\Delta \gamma A(\zeta _{1})\eta -U_{%
\mathrm{el}}(\zeta ),
\end{equation*}%
where $A(\zeta )$ denotes the (projected) contact area at the magnification $%
\zeta $, and $A(\zeta _{1})\eta $ is the real contact area, which is larger
than the projected contact area $A(\zeta _{1})$, i.e. $\eta \geq 1$ (e.g. if
the rigid solid is rough and the elastic solid has a flat surface $\eta >1$,
see Ref. \cite{P2} for an expression for $\eta $). $U_{\mathrm{el}}(\zeta )$
is the elastic energy stored at the interface due to the elastic deformation
of the solids on length scale shorter than $\lambda =L/\zeta $, necessary in
order to bring the solids into adhesive contact.

The area of apparent contact (projected on the $xy$-plane) at the
magnification $\zeta$, $A(\zeta)$, normalized by the nominal contact area $%
A_0$, can be obtained from 
\begin{equation*}
{\frac{A(\zeta)}{A_0}} = \int_{-\sigma_{\mathrm{a}}(\zeta)}^\infty d\sigma \
P(\sigma, \zeta)
\end{equation*}

Finally, we note that the effective interfacial energy to be used in the JKR
expression for the pull-off force (\ref{1}) is the macroscopic effective
interfacial energy corresponding to the magnification $\zeta =1$ (here we
assume that the reference length $L$ is of order the diameter of the JKR
contact region). Thus in the numerical results presented in Sec. 4 we only
study the area of contact $A(\zeta _{1})$ and the macroscopic interfacial
energy $\gamma _{\mathrm{eff}}=\gamma _{\mathrm{eff}}(1)$, which satisfies 
\begin{equation*}
\gamma _{\mathrm{eff}}A_{0}=\Delta \gamma A(\zeta _{1})\eta -U_{\mathrm{el}%
}(1)
\end{equation*}

\begin{figure}[tbp]
\includegraphics[width=0.45\textwidth]{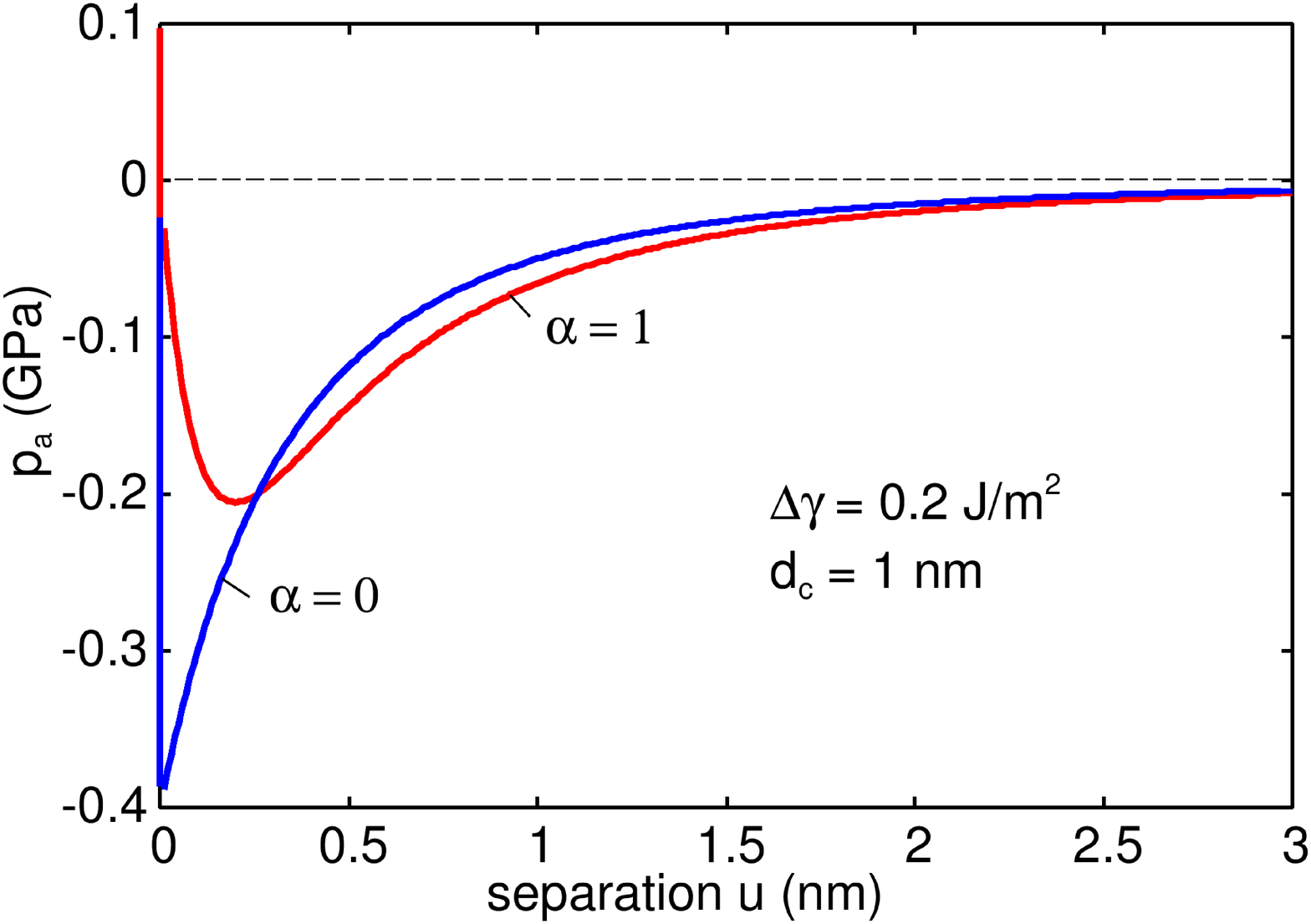}
\caption{The adhesive pressure for $n=3$ and $m=9$ and $\protect\alpha =1$
(red line) and $\protect\alpha =0$ (blue line). In the calculation we
assumed $\Delta \protect\gamma =0.2\ \mathrm{J/m^{2}}$ and $d_{\mathrm{c}%
}=1\ \mathrm{nm}$. }
\label{adhesivepressure.eps}
\end{figure}

\vskip 0.3cm \textbf{3.2 DMT-limit}

Let $p_{\mathrm{N}}=F_{\mathrm{N}}/A_{0}$ be the applied pressure (which can
be both positive and negative). In the DMT-limit one assumes that the elastic
deformation of the solids is the same as in the absence of an adhesive
interaction, except that the external load $F_{\mathrm{N}}$ is replaced with
an effective load. The latter contains the contribution to the normal force
from the adhesive force acting in the
non-contact interfacial surface area: $F_{0}=F_{\mathrm{N}}+F_{\mathrm{ad}}$%
. If we divide this equation by the nominal contact area $A_{0}$ we get 
\begin{equation*}
p_{0}=p_{\mathrm{N}}+p_{\mathrm{ad}}.
\end{equation*}%
The adhesive pressure 
\begin{equation}
p_{\mathrm{ad}}={\frac{1}{A_{0}}}\int_{\mathrm{n.c.}}d^{2}x\ p_{\mathrm{a}%
}(u(\mathbf{x}))  \label{2}
\end{equation}%
where $p_{\mathrm{a}}(u)$ is the interaction force per unit area when two
flat surfaces are separated by the distance $u$. In (\ref{2}) the integral
is over the non-contact (n.c.) area. In the study below we assume that the
is an attractive force per unit area between the surfaces given by ($u\geq 0$%
, see e.g. Fig. \ref{adhesivepressure.eps}): 
\begin{equation}
p_{\mathrm{a}}=B\left[ \left( {\frac{d_{\mathrm{c}}}{u+d_{\mathrm{c}}}}%
\right) ^{n}-\alpha \left( {\frac{d_{\mathrm{c}}}{u+d_{\mathrm{c}}}}\right)
^{m}\right] ,  \label{interacion.law}
\end{equation}%
where the cut-off $d_{\mathrm{c}}$ is a typical bond length and $\alpha $ a
number, which we take to be either 0 or 1 below. The parameter $B$ is
determined by the work of adhesion (per unit surface area): 
\begin{equation*}
\int_{0}^{\infty }du\ p_{\mathrm{ad}}(u)=Bd_{\mathrm{c}}{\frac{%
(m-1)-(n-1)\alpha }{(m-1)(n-1)}}=\Delta \gamma ,
\end{equation*}%
so that 
\begin{equation*}
B={\frac{\Delta \gamma }{d_{\mathrm{c}}}}{\frac{(m-1)(n-1)}{(m-1)-\alpha
(n-1)}}.
\end{equation*}%
If $P(u)$ denotes the distribution of interfacial separations then we can
also write (\ref{2}) as 
\begin{equation*}
p_{\mathrm{ad}}=\int_{0^{+}}^{\infty } du\ p_{\mathrm{a}}(u)P(u).
\end{equation*}%
In Ref. \cite{Carlos} we have derived an expression for $P(u)$ using the
Persson contact mechanics theory. In the numerical results presented below
we have used the expression for $P(u)$ given by Eq. (17) in Ref. \cite%
{Carlos} and below.

The effective interfacial energy can in the DMT-limit be calculated using 
\begin{equation*}
\gamma _{\mathrm{eff}}=\int_{0^{+}}^{\infty }du\ \phi (u)P(u)+\Delta \gamma {A \over A_0}-{\frac{U_{%
\mathrm{el}}}{A_{0}}}
\end{equation*}%
where $A=A_{\rm r}$ is the (repulsive) contact area and
where $\phi (u)$ is the interaction potential per unit surface area for flat
surfaces separated by the distance $u$ and given by 
\begin{equation*}
\phi (u)=\int_{u}^{\infty }du\ p_{\mathrm{a}}(u).
\end{equation*}%
Thus in the present case 
\begin{equation*}
\phi (u)={\frac{Bd_{\mathrm{c}}}{n-1}}\left( {\frac{d_{\mathrm{c}}}{u+d_{%
\mathrm{c}}}}\right) ^{n-1}-{\frac{Bd_{\mathrm{c}}\alpha }{m-1}}\left( {%
\frac{d_{\mathrm{c}}}{u+d_{\mathrm{c}}}}\right) ^{m-1}.
\end{equation*}%
For $u=0$ an infinite hard wall occurs and we define the (repulse) contact
area $A_{\mathrm{r}}$ when the surface separation $u=0$. We also define the
attractive contact area $A_{\mathrm{a}}$ when the surface separation $0<u<d_{%
\mathrm{c}}$, but this definition is somewhat arbitrary and another
definition was used in Ref. \cite{mark}. In the calculations below we use $%
n=3$ and $m=9$ and $\alpha =0$ (Sec. 4) and $\alpha =1$ (Sec. 5). The
interaction pressure for these two cases are shown in Fig. \ref%
{adhesivepressure.eps}.

The probability distribution of interfacial separations $P(u)$ can be
calculated as follows: We define $u_{1}(\zeta )$ to be the (average) height
separating the surfaces which appear to come into contact when the
magnification decreases from $\zeta $ to $\zeta -\Delta \zeta $, where $%
\Delta \zeta $ is a small (infinitesimal) change in the magnification. $%
u_{1}(\zeta )$ is a monotonically decreasing function of $\zeta $, and can
be calculated from the average interfacial separation $\bar{u}(\zeta )$ and
the contact area $A(\zeta )$ using (see Ref. \cite{YP}) 
\begin{equation*}
u_{1}(\zeta )=\bar{u}(\zeta )+\bar{u}^{\prime }(\zeta )A(\zeta )/A^{\prime
}(\zeta ).
\end{equation*}%
The equation for the average interfacial separation $\bar{u}(\zeta )$ is
given in Ref. \cite{YP}. The (apparent) relative contact area $A(\zeta
)/A_{0}$ at the magnification $\zeta $ is given by 
\begin{equation*}
{\frac{A(\zeta )}{A_{0}}}=\mathrm{erf}\left( \frac{p_{0}}{2G(\zeta )^{1/2}}%
\right) ,
\end{equation*}%
where 
\begin{equation*}
G(\zeta )={\frac{\pi }{4}}\left( {\frac{E}{1-\nu ^{2}}}\right)
^{2}\int_{q_{0}}^{\zeta q_{0}}dqq^{3}C(q),
\end{equation*}%
where $C(q)$ is the surface roughness power spectrum. In what follows we
will denote this contact area as the \textit{repulsive} contact area $A_{%
\mathrm{r}}$ since the normal stress is repulsive within this area. We also
define an \textit{attractive} contact area $A_{\mathrm{a}}$ as the surface
area where the surface separation $0<u < d_{\mathrm{c}}$; in this surface
separation interval the wall-wall interaction is attractive. The cut-off
length $d_{\mathrm{c}}$ is quite arbitrary and in Ref. \cite{mark} another
cut-off length (of order $d_{\mathrm{c}}$) was used to define the attractive
contact area.

The probability distribution $P(u)$ can be written as\cite{Carlos} 
\begin{equation*}
P(u) \approx \frac{1}{A_0} \int d\zeta \ [-A^{\prime }(\zeta)] \frac{1}{%
\left ( 2\pi h^2_{\mathrm{rms}}(\zeta) \right )^{1/2}}
\end{equation*}
\begin{equation*}
\times \left [ \mathrm{exp} \left ( -\frac{(u-u_1(\zeta))^2}{2 h^2_{\mathrm{%
rms}}(\zeta)}\right ) + \mathrm{exp} \left ( - \frac{(u+u_1(\zeta))^2}{2
h^2_{\mathrm{rms}}(\zeta)}\right )\right ] ,
\end{equation*}
where $h^2_{\mathrm{rms}}(\zeta)$ is the mean of the square of the surface
roughness amplitude including only roughness components with the wavevector $%
q>q_0\zeta$, and given by 
\begin{equation*}
h^2_{\mathrm{rms}}(\zeta) = \int_{q> q_0\zeta} d^2q \ C(q).
\end{equation*}

\vskip 0.3cm \textbf{3.3 Scale-dependent Tabor length $d_{\mathrm{T}}(q)$}

The contact between surfaces with roughness on many length scales involves
contact between asperities with many different radius of curvatures. Thus at
low magnifications we only observe long-wavelength roughness and the
asperity radius of curvature may be macroscopic, e.g., $\sim 1\ \mathrm{mm}$
or more. At high magnification, nanoscale roughness will be observed
involving asperities which may have radius of curvature in the nm range.
Thus adhesion at long length scale may appear JKR-like while at short enough
length scale the adhesion may appear DMT-like. One can define a
magnification or length-scale dependent Tabor length $d_{\mathrm{T}}(\zeta )$
($\zeta =q/q_{0}$), in the following way: If we include only roughness
components with wavevector $q<\zeta q_{0}$ the mean summit asperity
curvature is\cite{Nyak} 
\begin{equation*}
{\frac{1}{R^{2}(\zeta )}}={\frac{16}{3\pi }}\int_{q_{0}}^{\zeta q_{0}}dq\
q^{5}C(q)
\end{equation*}%
We define 
\begin{equation*}
d_{\mathrm{T}}(\zeta )=\left( {\frac{R(\zeta )[\gamma _{\mathrm{eff}}(\zeta
)]^{2}}{E_{\mathrm{r}}^{2}}}\right) ^{1/3}
\end{equation*}%
If $d_{\mathrm{T}}(\zeta )<<d_{\mathrm{c}}$ the contact at the magnification 
$\zeta =q/q_{0}$, will appear DMT-like while if $d_{\mathrm{T}}(\zeta )>>d_{%
\mathrm{c}}$ the contact will appear JKR-like. In what follows we will
sometimes denote $d_{\mathrm{T}}(\zeta )$ with $d_{\mathrm{T}}(q)$ ($q=\zeta
q_{0}$).

\vskip 0.3cm \textbf{4 Theory: numerical results}

We now present numerical results which illustrates the two adhesion theories
presented above. The JKR-like theory has been studied before (see Ref. \cite%
{P2}) so we focus mainly on the DMT-like theory. In the calculations we vary 
$\Delta \gamma $ and $n$, but we always use the cut-off $d_{\mathrm{c}}=0.4\ 
\mathrm{nm}$ and $\alpha =0$ unless otherwise stated.

\begin{figure}[tbp]
\includegraphics[width=0.45\textwidth]{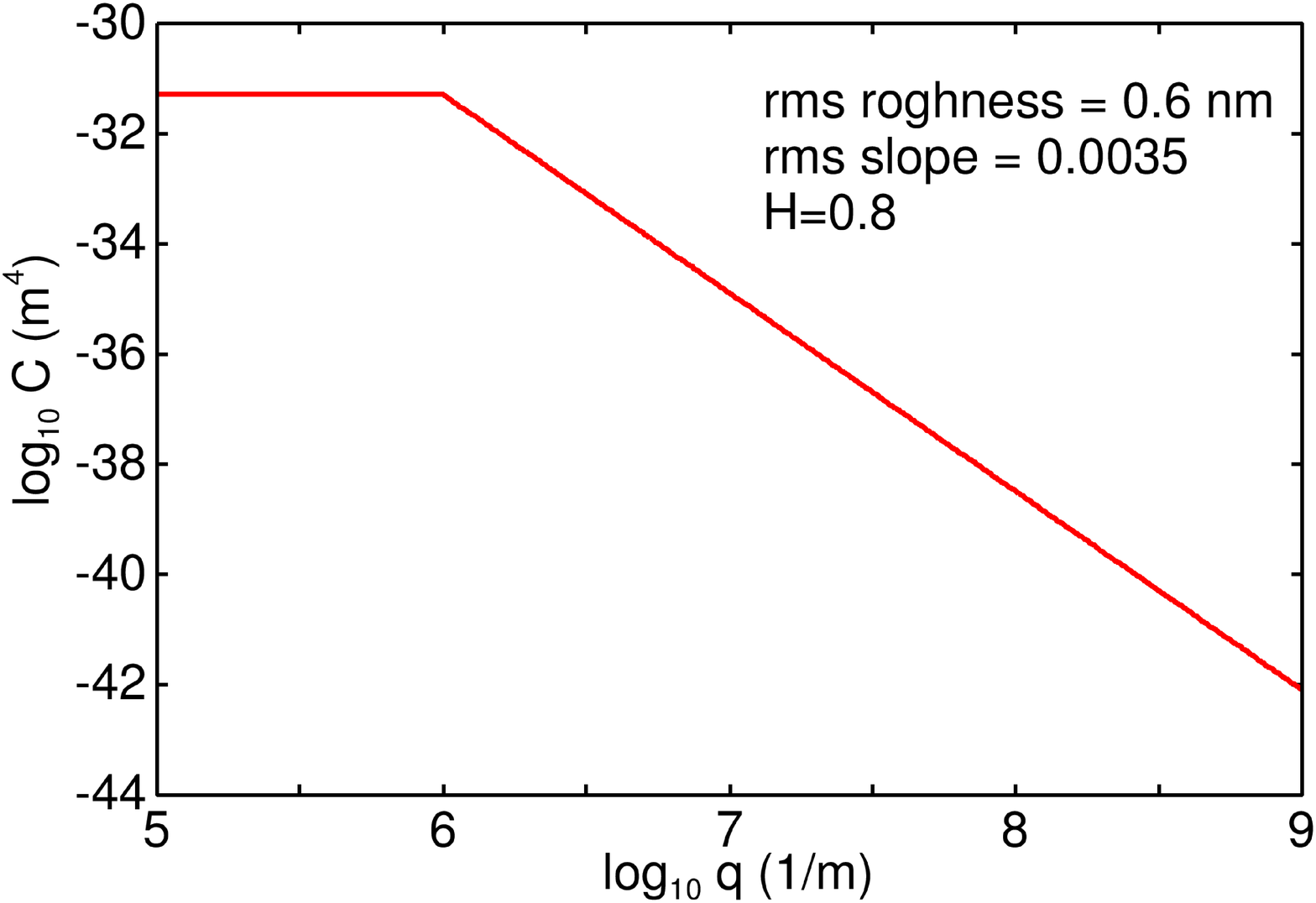}
\caption{The surface roughness power spectrum $C(q)$ as a function of the
wavevector $q$ ($\mathrm{log}_{10}-\mathrm{log}_{10}$ scale), used in the
present calculations. The power spectrum corresponds to a surface with the
rms roughness amplitude $0.6 \ \mathrm{nm}$, the rms slope $0.0035$ and the
Hurst exponent $H=0.8$. }
\label{1logq.2logC.eps}
\end{figure}

\begin{figure}[tbp]
\includegraphics[width=0.45\textwidth]{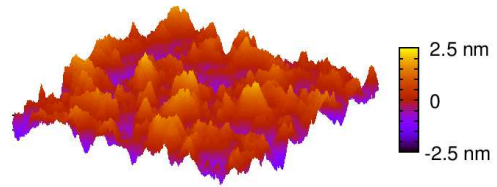}
\caption{Surface topography of one realization of a surface with the surface
roughness power spectrum shown in Fig. \protect\ref{1logq.2logC.eps}. The
difference between the lowest and highest point is about $5 \ \mathrm{nm}$,
i.e. about 10 times higher than the rms roughness $0.6 \ \mathrm{nm}$ (see Appendix A in \cite{P3}).}
\label{snapshot.pic.eps}
\end{figure}

\begin{figure}[tbp]
\includegraphics[width=0.45\textwidth]{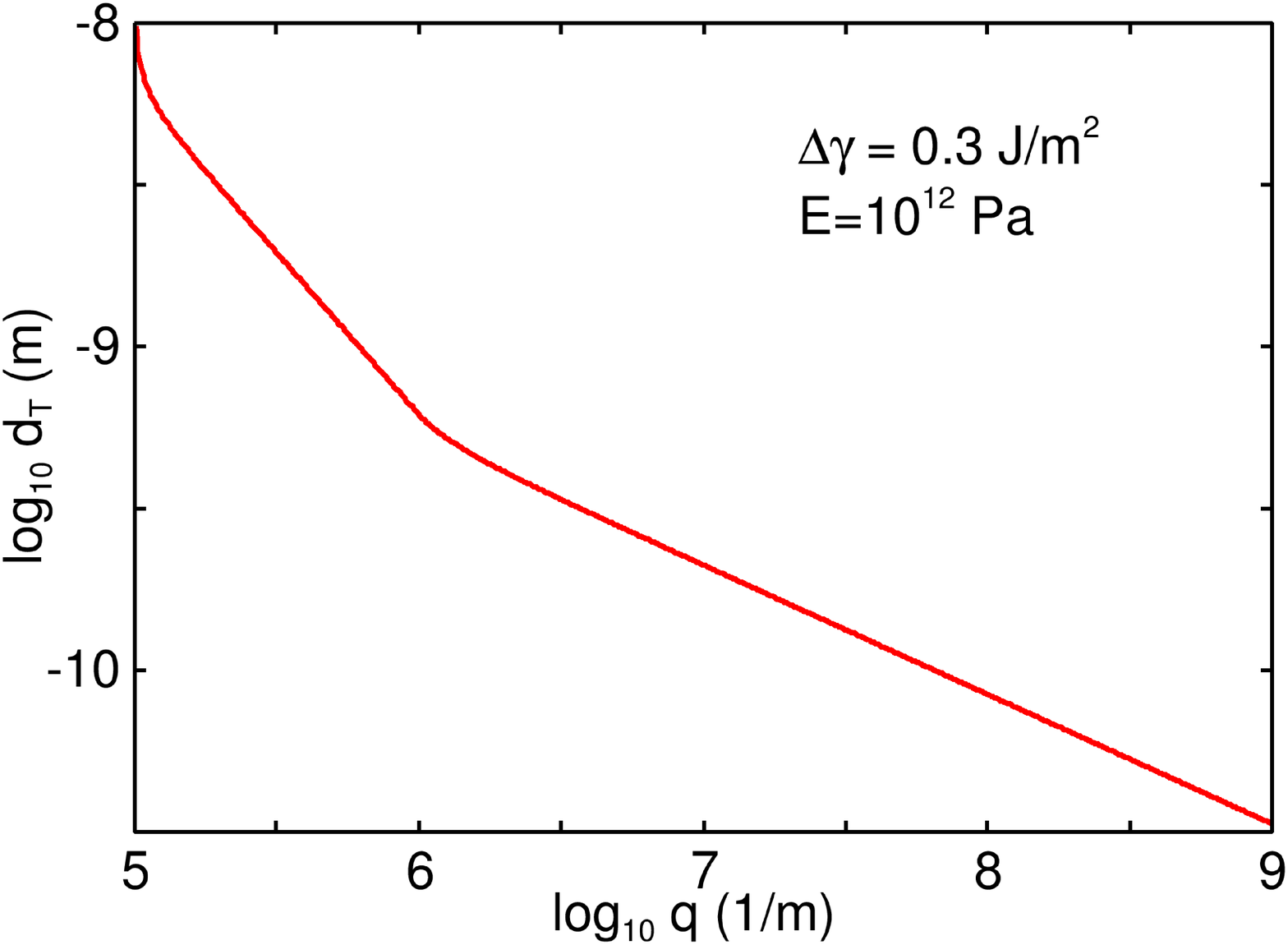}
\caption{The Tabor length parameter $d_{\mathrm{T}}$ as a function of the
wavevector ($\mathrm{log}_{10}-\mathrm{log}_{10}$ scale). For the surface
with the power spectrum given in Fig. \protect\ref{1logq.2logC.eps} and with
the elastic modulus $E=10^{12} \ \mathrm{Pa}$ ($\nu=0.5$) and work of adhesion $\Delta 
\protect\gamma = 0.3 \ \mathrm{J/m^2}$. }
\label{1logq.2log.Tabor.Length.eps}
\end{figure}

\vskip 0.3cm \textbf{4.1 Surface roughness power spectrum $C(q)$ and Tabor
length $d_{\mathrm{T}} (q)$}

In Fig. \ref{1logq.2logC.eps} we show the surface roughness power spectrum $%
C(q)$ (PSD) as a function of the wavevector $q$ ($\mathrm{log}_{10}-\mathrm{%
log}_{10}$ scale), used in the present calculations. The power spectrum
corresponds to a surface with the rms roughness amplitude $0.6\ \mathrm{nm}$%
, the rms slope $0.0035$ and the Hurst exponent $H=0.8$. Fig. \ref%
{snapshot.pic.eps} shows the surface topography of one realization of a
randomly rough surface with the surface roughness power spectrum shown in
Fig. \ref{1logq.2logC.eps}. The difference between the lowest and highest
point is about $5\ \mathrm{nm}$, i.e. about 10 times higher than the rms
roughness $0.6\ \mathrm{nm}$.

In the calculations below we use the Young's modulus $E=10^{12} \ \mathrm{Pa}
$, Poisson ration $\nu=0.5$ and the work of adhesion $\Delta \gamma =
0.1-0.4 \ \mathrm{J/m^2}$. Fig \ref{1logq.2log.Tabor.Length.eps} shows the
Tabor length parameter $d_{\mathrm{T}}$ as a function of the wavevector ($%
\mathrm{log}_{10}-\mathrm{log}_{10}$ scale), for $\Delta
\gamma = 0.3 \ \mathrm{J/m^2}$. Note that the contact mechanics is DMT-like
for short length scales (or large wavevectors) with $d_{\mathrm{T}} < d_{%
\mathrm{c}} = 0.4 \ \mathrm{nm}$, while it is JKR-like for long length
scales (small wave vectors).

\begin{figure}[tbp]
\includegraphics[width=0.45\textwidth]{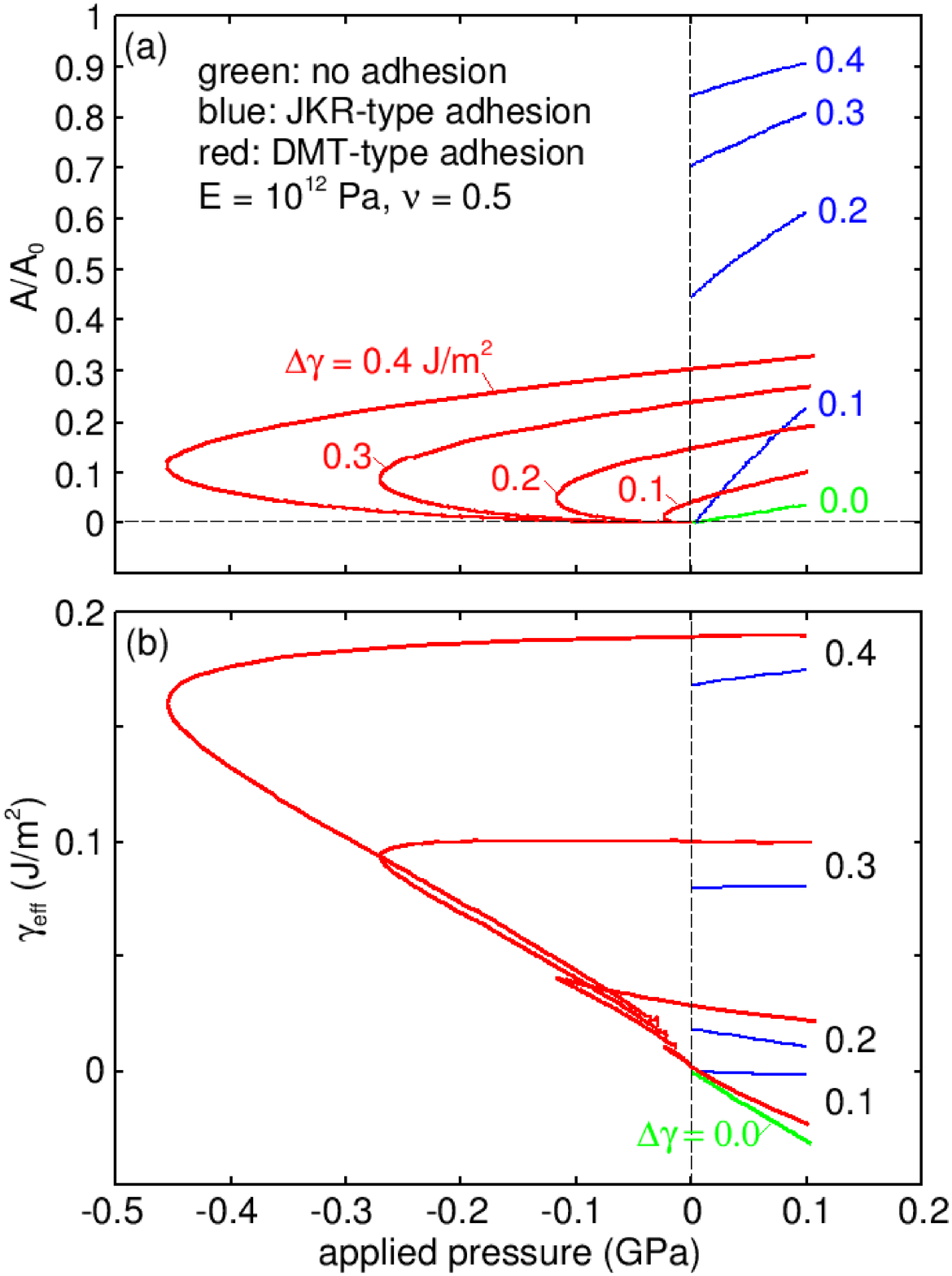}
\caption{The normalized (projected) area of contact $A/A_0$ and the
effective interfacial energy $\protect\gamma_{\mathrm{eff}} = [E_{\mathrm{ad}%
}-U_{\mathrm{el}}]/A_0$ (where $E_{\mathrm{ad}}$ is the (attractive) Van der
Waals interaction energy and $U_{\mathrm{el}}$ the (repulsive) elastic
deformation energy) as a function of the applied (nominal) pressure $p_{\mathrm{N}}$
acting on the block. In the DMT-like theory (red curve) $A=A_{\mathrm{r}%
} $ is the repulsive contact area while in the JKR-like theory (blue
curves) $A$ is the total contact area (which has both an attractive and a
repulsive part). Results are shown for the work of adhesion $\Delta \protect%
\gamma = 0.0$ (green curve), $0.1$, $0.2$, $0.3$ and $0.4 \ \mathrm{J/m^2}$.
The red and blue lines correspond to DMT-like and JKR-like approximations, respectively.
The elastic solid Young's modulus $E=10^{12} \ \mathrm{Pa}$ and
Poisson number $\protect\nu = 0.5$. }
\label{combine.p.A.p.gamma.eps}
\end{figure}

\begin{figure}[tbp]
\includegraphics[width=0.45\textwidth]{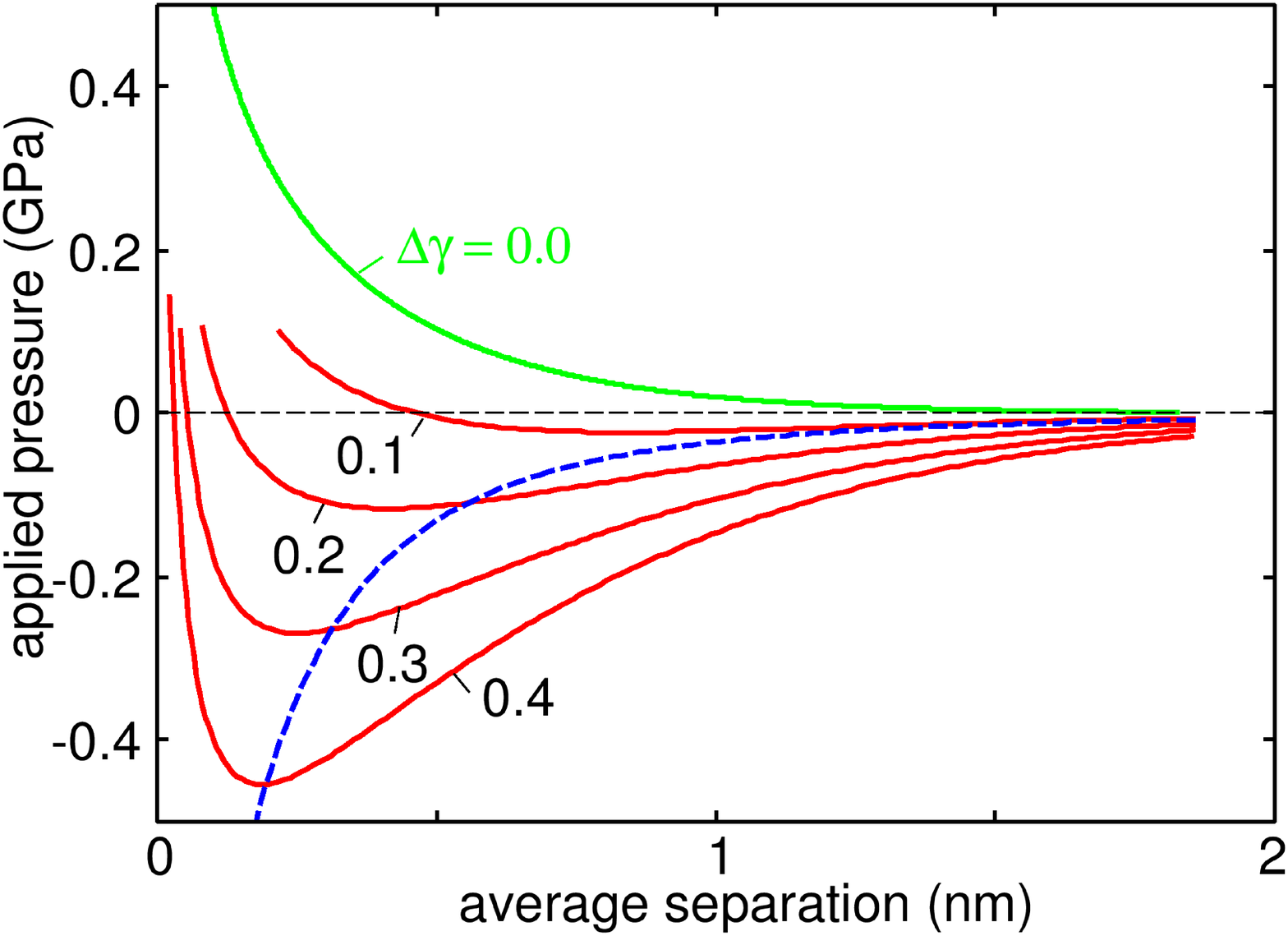}
\caption{The applied pressure as a function of the average separation for
the work of adhesion $\Delta \protect\gamma = 0.0$ (green curve), $0.1$, $%
0.2 $, $0.3$ and $0.4 \ \mathrm{J/m^2}$. The blue dashed curve is the Van
der Waals interaction force per unit area $p_{\mathrm{a}}=B[d_{\mathrm{c}%
}/(u+d_{\mathrm{c}})]^3$, where $d_{\mathrm{c}}=0.4 \ \mathrm{nm}$ and $u$
the distance from the hard wall. The parameter $B$ is chosen to reproduce
the given work of adhesion for flat surfaces. }
\label{interpot.eps}
\end{figure}

\vskip 0.3cm \textbf{4.2 Results for different work of adhesion $\Delta
\gamma$}

Fig. \ref{combine.p.A.p.gamma.eps} shows the normalized (projected) area of
contact $A/A_0 $ and the effective interfacial energy $\gamma_{\mathrm{eff}}
= [E_{\mathrm{ad}}-U_{\mathrm{el}}]/A_0$ (where $E_{\mathrm{ad}}$ is the
(attractive) Van der Waals interaction energy and $U_{\mathrm{el}}$ the
(repulsive) elastic deformation energy) as a function of the nominal applied
pressure $p_{\mathrm{N}}$ acting on the block. In the DMT-like theory
(red curve) $A=A_{\mathrm{r}}$ is the repulsive contact area while in the
JKR-like theory (blue curves) $A$ is the total contact area (which has
both an attractive and a repulsive part). Results are shown for the work of
adhesion $\Delta \gamma = 0.0$ (green curve), $0.1$, $0.2$, $0.3$ and $0.4 \ 
\mathrm{J/m^2}$. The red and blue lines correspond to DMT-like and JKR-like
approximations, respectively. Note that the area of contact is about a factor of $3$
larger in the JKR-like approximation as compared to the DMT-like
approximation. This is consistent with the results for adhesion of sphere on
flat (see Sec. 2) where the JKR theory predict about 2 times larger contact
area than the DMT theory. On the other hand Fig. \ref{combine.p.A.p.gamma.eps}%
(b) shows that the effective interfacial binding energies are similar, which
is also consistent with the results of Sec. 2. The effective work of
adhesion to be used in macroscopic adhesion applications, i.e., the pull-off
of a ball from a flat (Sec. 2) is $\gamma_{\mathrm{eff}}$ for the applied
pressure $p_{\mathrm{N}}=0$, and in all cases in Fig. \ref%
{combine.p.A.p.gamma.eps} $\gamma_{\mathrm{eff}}(p_{\mathrm{N}}=0)$ is less
than half of the work of adhesion $\Delta \gamma$ for smooth surfaces.

Fig. \ref{combine.p.A.p.gamma.eps}(b) shows that for $\Delta \gamma = 0.1 \ 
\mathrm{J/m^2}$ in the JKR-limit the effective interfacial binding energy,
and hence also the pull-off force, vanish. Nevertheless, Fig. \ref%
{combine.p.A.p.gamma.eps}(a) shows that in the JKR-limit the contact area as
a function of $p_{\mathrm{N}}$ increases much faster with increasing $p_{%
\mathrm{N}}$ than in the absence of adhesion (green line), i.e., even if no
adhesion manifests itself during pull-off, the contact area and hence other
properties like the friction force, may be strongly enhanced by the adhesive
interaction. In the DMT-limit the effective interfacial binding energy is
always non-zero if the wall-wall interaction does not vanish beyond some fix
wall-wall separation. This is easy to understand since when the wall-wall
separation is larger than the highest asperity the solid walls will only
interact with the long-ranged attractive wall-wall potential and increasing
the separation to infinity will always require a finite amount of work
making $\gamma_{\mathrm{eff}}(p_{\mathrm{N}}=0)$ always non-zero in the
DMT-limit.

Let us now discuss the slopes (with increasing $p_{\mathrm{N}}$) of the $%
\gamma_{\mathrm{eff}} (p_{\mathrm{N}})$ curves in Fig. \ref%
{combine.p.A.p.gamma.eps}(b) for $p_{\mathrm{N}} = 0$. As pointed out in Sec.
2, in an exact treatment, as a function of the external load $p_{\mathrm{N}}$
the total energy $-A_0 \gamma_{\mathrm{eff}}=-E_{\mathrm{ad}}+U_{\mathrm{el}%
} $ must have a minimum at $p_{\mathrm{N}} = 0$. However, the theories
described above are not exact, and are not based on a treatment which
minimize the total energy, but rather focus on the force (or stress) (in the
DMT-like model) or on a combined energy and stress treatment (in the
JKR-like model). This is the reason for why the slope of the $\gamma_{%
\mathrm{eff}} (p_{\mathrm{N}})$ curves for large $\Delta \gamma$ is positive
rather than negative. However, the slope is rather small compared to the
(absolute value of) the slope for the non-adhesive interaction (green
curve). In addition, the surface we use has a Tabor length with $d_{\mathrm{T%
}} (q)<< d_{\mathrm{c}}$ for large $q$ and $d_{\mathrm{T}} (q)>> d_{\mathrm{c%
}}$ for small $q$ so strictly speaking neither the JKR-limit or the
DMT-limit is correct or valid. For other surfaces which have $d_{\mathrm{T}}
(q)<< d_{\mathrm{c}}$ or $d_{\mathrm{T}} (q)>> d_{\mathrm{c}}$ for all $q$,
the JKR-like and DMT-like theories may be more accurate and the slope of the 
$\gamma_{\mathrm{eff}} (p_{\mathrm{N}})$ curve negative.

Fig. \ref{interpot.eps} shows the applied pressure as a function of the
average separation for the work of adhesion $\Delta \gamma =0.0$ (green
curve), $0.1$, $0.2$, $0.3$ and $0.4\ \mathrm{J/m^{2}}$. The blue dashed
curve is the Van der Waals interaction force per unit area $p_{\mathrm{a}%
}=B[d_{\mathrm{c}}/(u+d_{\mathrm{c}})]^{3}$, where $d=0.4\ \mathrm{nm}$ and $%
u$ the distance from the hard wall. The parameter $B$ is chosen to reproduce
the given work of adhesion for flat surfaces. Note that the attractive
interaction between the walls is already strong at distances where the 
flat surfaces negligible wall-wall interaction would occur (as described by
the blue dashed line). This is of course due to adhesive interaction
involving high asperities, which prevail even when the average wall-wall
separation is relative large. For the case of no adhesion (green curve) the
wall-wall interaction is purely repulsive as the asperities get compressed
on decreasing the wall-wall separation. Asymptotically (large separation) this repulsive
interaction is exponential $p_{\mathrm{N}}\sim \mathrm{exp}(-u/u_{0})$ where
the reference length $u_{0}$ is of order the rms surface roughness amplitude.

\begin{figure}[tbp]
\includegraphics[width=0.45\textwidth]{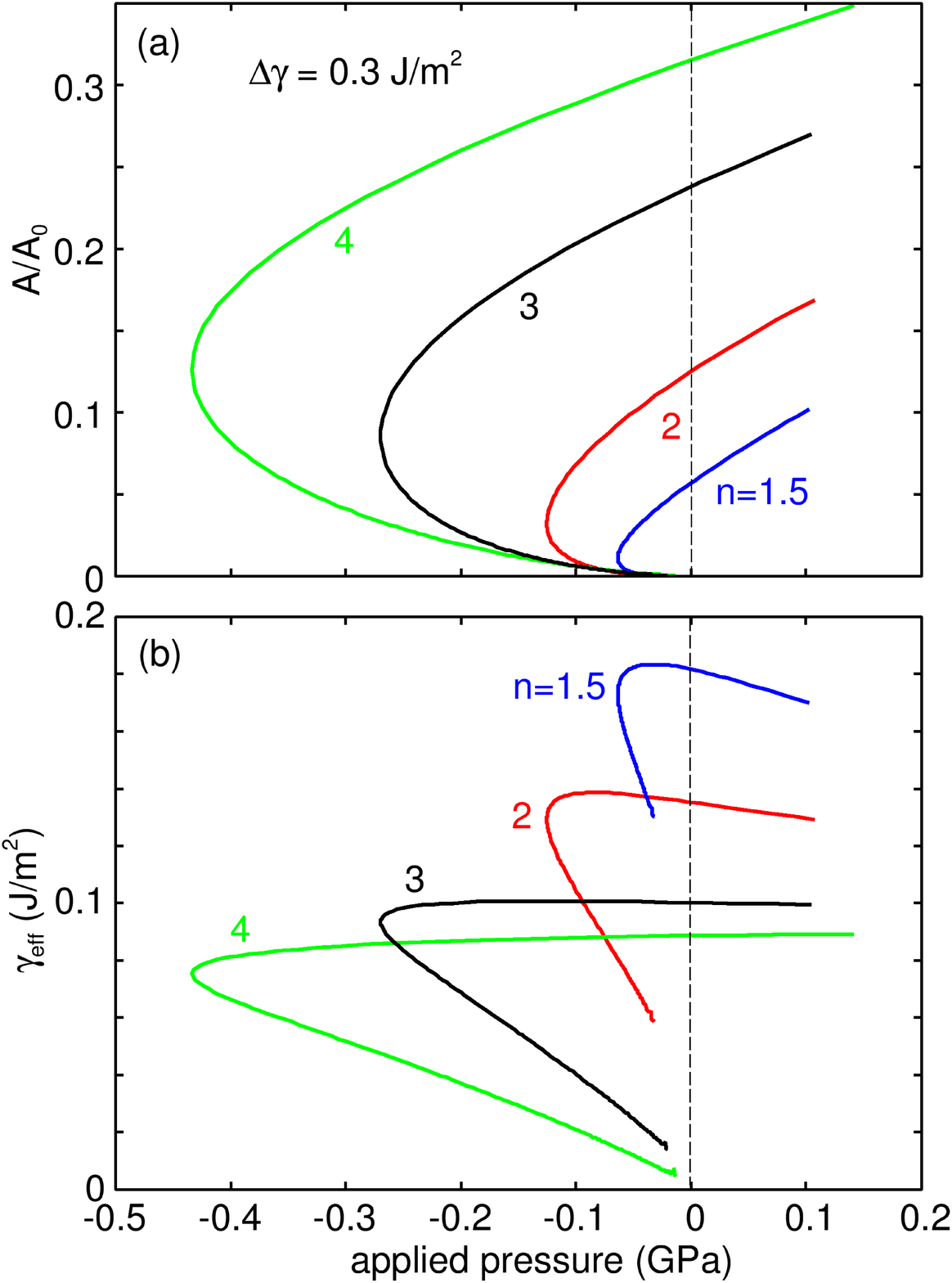}
\caption{The normalized (projected) repulsive area of contact $A/A_{0}$ and
the effective interfacial energy $\protect\gamma _{\mathrm{eff}}=[E_{\mathrm{%
ad}}-U_{\mathrm{el}}]/A_{0}$ (where $E_{\mathrm{ad}}$ is the (attractive)
Van der Waals interaction energy and $U_{\mathrm{el}}$ the (repulsive)
elastic deformation energy) as a function of the nominal pressure $p_{%
\mathrm{N}}$ acting on the block. Results are shown for the work of adhesion 
$\Delta \protect\gamma =0.3\ \mathrm{J/m^{2}}$ and the interaction force
index $n=1.5$, 2, 3 and $4$ ($p_{\mathrm{a}}=B[d_{\mathrm{c}}/(u+d_{\mathrm{c%
}})]^{n}$). From the DMT-like approximation (see text). The elastic solid
Young's modulus $E=10^{12}\ \mathrm{Pa}$ and Poisson number $\protect\nu %
=0.5 $. }
\label{1p.2gammaeff.and.A.n.eq.1.5.to.4.eps}
\end{figure}

\begin{figure}[tbp]
\includegraphics[width=0.45\textwidth]{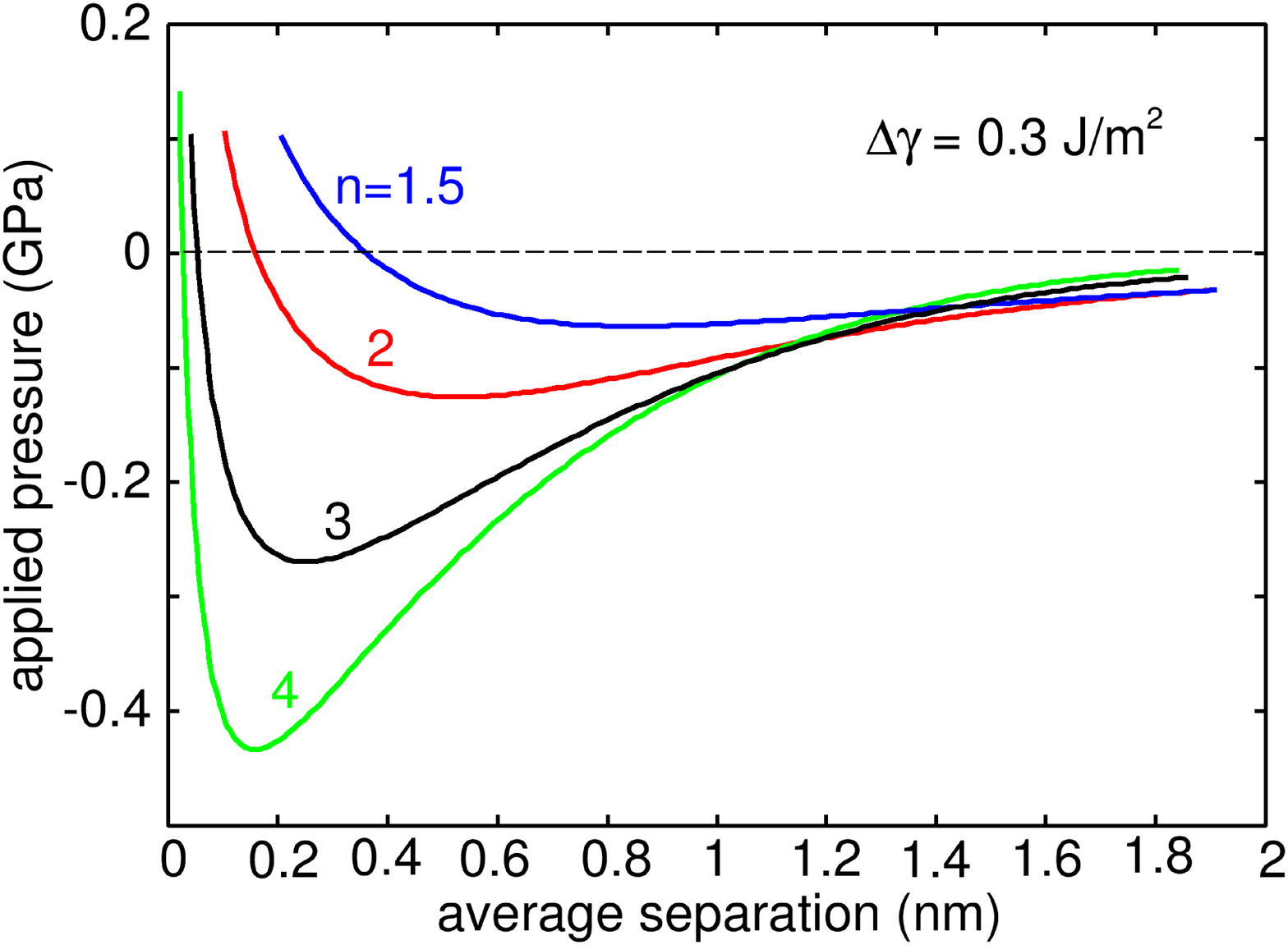}
\caption{The applied pressure $p_{\mathrm{N}}$ as a function of the average
separation for the work of adhesion $\Delta \protect\gamma =0.3\ \mathrm{%
J/m^{2}}$ and the interaction force index $n=1.5$, 2, 3 and $4$ ($p_{\mathrm{%
a}}=B[d_{\mathrm{c}}/(u+d_{\mathrm{c}})]^{n}$). From the DMT-like
approximation (see text). }
\label{1u.2p.n.eq.1.5.to.4.eps}
\end{figure}

\begin{figure}[tbp]
\includegraphics[width=0.45\textwidth]{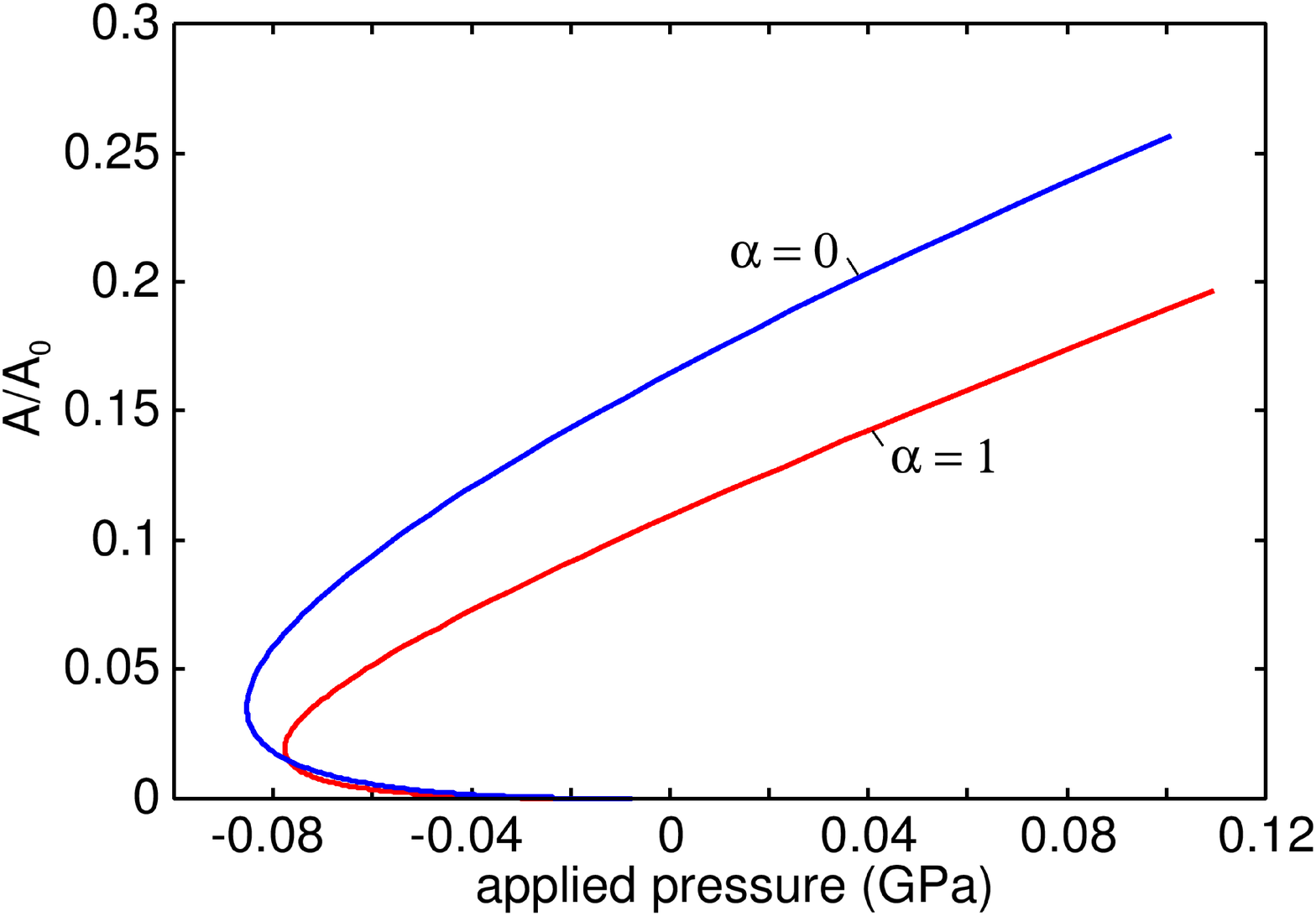}
\caption{The normalized (projected) repulsive area of contact $A_{\mathrm{r}%
}/A_{0}$ as a function of the nominal pressure $p_{\mathrm{N}}$ acting on
the block. Results are shown for the work of adhesion $\Delta \protect\gamma %
=0.2\ \mathrm{J/m^{2}}$ and the interaction index $n=3$ and $m=9$ and with $%
d_{\mathrm{c}}=1\ \mathrm{nm}$. Blue curve is with $\protect\alpha =0$ and
red curve with $\protect\alpha =1$. From the DMT-like approximation (see
text). }
\label{1pullp.2A.blue.first.eps}
\end{figure}

\begin{figure}[tbp]
\includegraphics[width=0.45\textwidth]{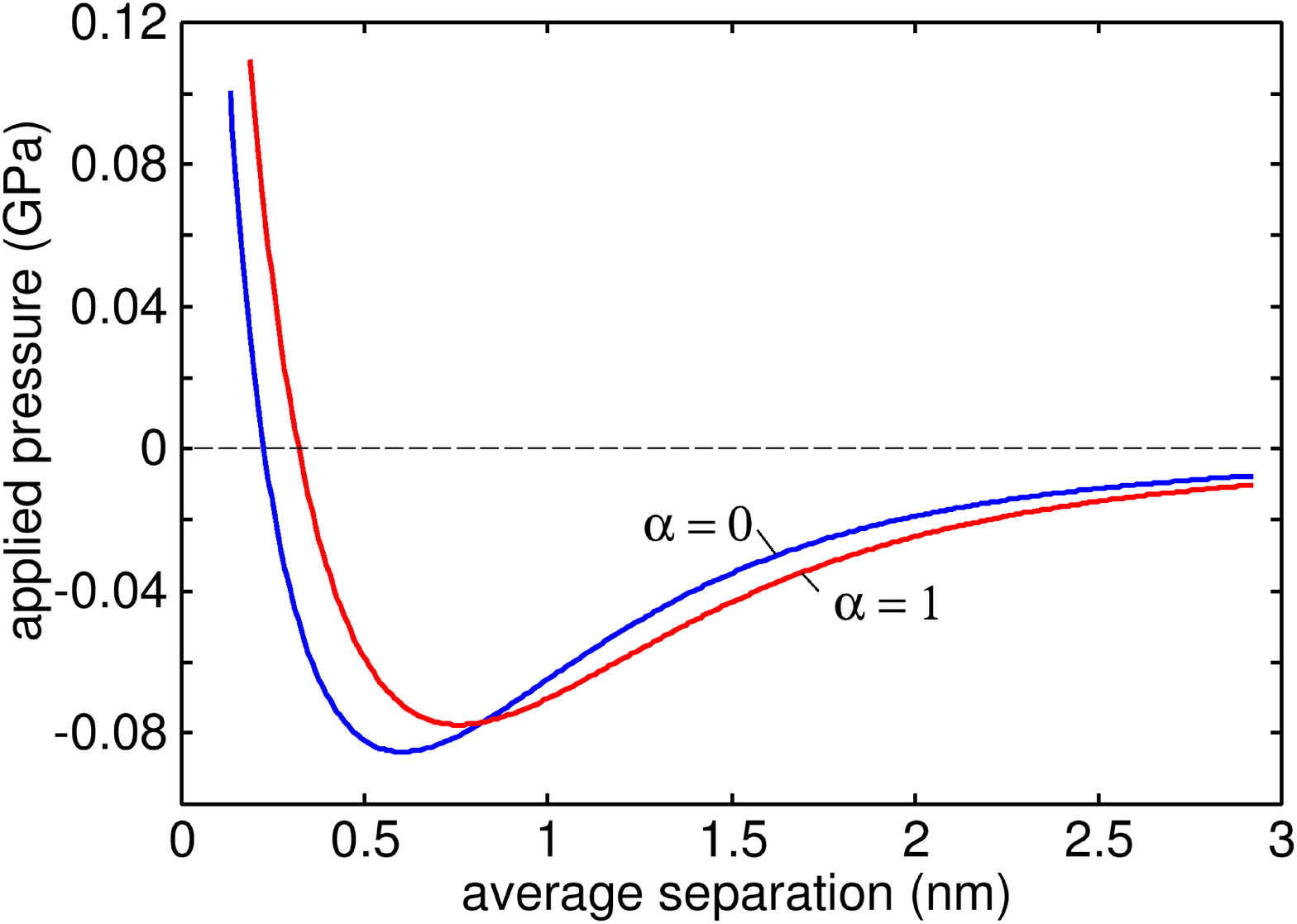}
\caption{The applied pressure $p_{\mathrm{N}}$ as a function of the average
separation for the work of adhesion $\Delta \protect\gamma =0.2\ \mathrm{%
J/m^{2}}$ and the interaction index $n=3$ and $m=9$ and with $d_{\mathrm{c}%
}=1\ \mathrm{nm}$. Blue curve is with $\protect\alpha =0$ and red curve with 
$\protect\alpha =1$. From the DMT-like approximation (see text). }
\label{1separation.2pN.blue.first.eps}
\end{figure}

\vskip 0.3cm \textbf{4.3 Results for different interaction potential
exponent $n$ and factor $\alpha$}

Fig. \ref{1p.2gammaeff.and.A.n.eq.1.5.to.4.eps} shows the normalized
(projected) repulsive area of contact $A_{\mathrm{r}}/A_{0}$ and the
effective interfacial energy $\gamma _{\mathrm{eff}}=[E_{\mathrm{ad}}-U_{%
\mathrm{el}}]/A_{0}$ (where $E_{\mathrm{ad}}$ is the (attractive) Van der
Waals interaction energy and $U_{\mathrm{el}}$ the (repulsive) elastic
deformation energy) as a function of the nominal pressure acting on the
block. Results are shown for the work of adhesion $\Delta \gamma =0.3\ 
\mathrm{J/m^{2}}$ and the interaction force index $n=1.5\ ,2\ ,3$ and $4$ ($%
p_{\mathrm{a}}=B[d_{\mathrm{c}}/(u+d_{\mathrm{c}})]^{n}$, i.e. $\alpha =0$).
The results are for the DMT-like approximation. The elastic solid Young's
modulus $E=10^{12}\ \mathrm{Pa}$ and Poisson number $\nu =0.5$.

Fig. \ref{1p.2gammaeff.and.A.n.eq.1.5.to.4.eps} shows that as the interaction
becomes more short ranged ($n$ increases from 1.5 to 4) (at fixed work of
adhesion $\Delta \gamma$) the contact area increases while the effective
interfacial binding energy $\gamma_{\mathrm{eff}}$ decreases. The latter is
easy to understand: in the limiting case when $n\rightarrow 0$ the
interaction potential has infinite extend (and infinitesimal strength in
such a way that the work of adhesion $\Delta \gamma = 0.3 \ \mathrm{J/m^2}$)
and in this case $\gamma_{\mathrm{eff}}$ must equal $\Delta \gamma$. At the
same time due to the weak (infinitesimal) force the contact area at the load 
$p_{\mathrm{N}} = 0$ must vanish, which explain the behavior observed in Fig. %
\ref{1p.2gammaeff.and.A.n.eq.1.5.to.4.eps}(a).

Fig. \ref{1u.2p.n.eq.1.5.to.4.eps} shows the applied pressure as a function
of the average separation for the work of adhesion $\Delta \gamma =0.3\ 
\mathrm{J/m^{2}}$ and the interaction force index $n=1.5\ ,2\ ,3$ and $4$ ($%
p_{\mathrm{a}}=B[d_{\mathrm{c}}/(u+d_{\mathrm{c}})]^{n}$). The results are
for the DMT-like approximation. Note that when $n$ decreases the more
long-range the effective attraction but at the same time the smaller the
maximal attraction, which again reflect the fact that $\Delta \gamma $ is
kept fixed.

All the numerical results presented above was for the cut-off length $d_{%
\mathrm{c}}=0.4\ \mathrm{nm}$ and the repulsion factor $\alpha =0$. We now
consider the case $\alpha =1$ with $m=9$ (and $n=3$). We also use $d_{%
\mathrm{c}}=1.0\ \mathrm{nm}$. These are the same parameters we will use
when comparing the theory with exact numerical results in Sec. 5. We
consider a surface with the rms roughness $0.5\ \mathrm{nm}$, the roll-off
wavevector $q_{\mathrm{r}}=1.0\times 10^{6}\ \mathrm{m^{-1}}$ and the small
and large wavevector cut-off $q_{0}=2.5\times 10^{5}\ \mathrm{m^{-1}}$ and $%
q_{1}=3.2\times 10^{7}\ \mathrm{m^{-1}}$.

Fig. \ref{1pullp.2A.blue.first.eps} shows the normalized (projected)
repulsive contact area $A_{\mathrm{r}}/A_0$ as a function of the nominal
pressure $p_{\mathrm{N}}$ acting on the block. Results are shown for the
work of adhesion $\Delta \gamma = 0.2 \ \mathrm{J/m^2}$.
The blue curve is with $\alpha =0$ and red curve with $\alpha =1$.

Fig. \ref{1separation.2pN.blue.first.eps} shows the applied pressure $p_{%
\mathrm{N}}$ as a function of the average surface separation for the work of
adhesion $\Delta \gamma = 0.2 \ \mathrm{J/m^2}$. Again the blue curve is with 
$\alpha =0$ and red curve with $\alpha =1$.

\vskip0.3cm \textbf{5 Comparison of the DMT-like theory with exact numerical
results}

\begin{figure}[tbp]
\includegraphics[width=0.45\textwidth]{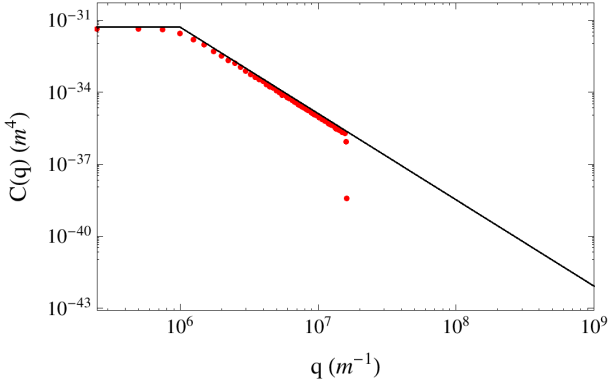}
\caption{Solid line: Power spectral density $C\left( q\right) $ as a
function of $q$ (solid black line). For an isotropic surface roughness with
cut-off frequency $q_{0}=q_{\mathrm{r}}/4$, root-mean-square roughness $h_{%
\mathrm{rms}}=0.6\ \mathrm{nm}$, and with self-affine regime in the
frequency range $q_{\mathrm{r}}=10^{6}\ \mathrm{m}^{-1}$ to $q_{1}=10^{3}q_{%
\mathrm{r}}$. The Hurst exponent is $H=0.8$. (dotted line): The PSD adopted
in the numerical calculations is truncated at $q_{1}=64q_{0}$, with 8
divisions at the smallest length scale ($q_{1}$), resulting in a $h_{\mathrm{%
rms}}=0.52\ \mathrm{nm}$ and mean square slope $0.00115$. }
\label{psd.eps}
\end{figure}

In this section we use the interaction potential (\ref{interacion.law}) with 
$n=3$, $m=9$ and $\alpha =1$ with $d_{\mathrm{c}}=1\ \mathrm{nm}$ 
(see Fig. \ref{adhesivepressure.eps} and Appendix A). The power spectral
density adopted in the numerical calculations (see Appendix \ref{appendix.1}
for the summary of the numerical model) is shown in Fig. \ref{psd.eps}.

\begin{figure}[tbp]
\includegraphics[width=0.45\textwidth]{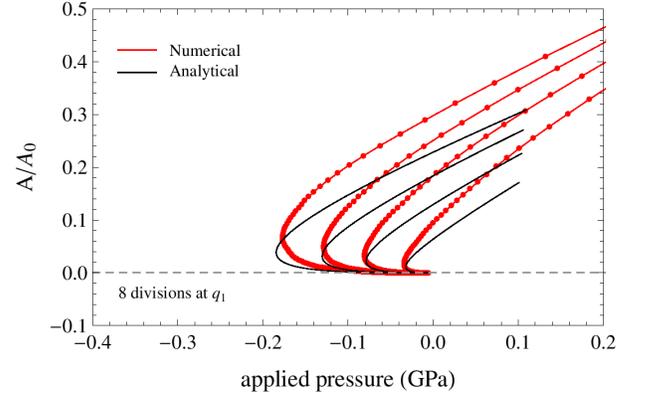}
\caption{Normalized (projected) area of repulsive contact $A_{\mathrm{r}}/A_{0}$ as a
function of the applied pressure $p_{\mathrm{N}}$, for different values of
work of adhesion $\Delta \protect\gamma =0.1,\ 0.2,\ 0.3,\ 0.4\ \mathrm{%
J/m^{2}}$. For an elastic solid with $E_{\mathrm{r}}=1.33\times 10^{12}\ \mathrm{%
GPa}$ and for the surface roughness of Fig. \protect\ref{psd.eps}.}
\label{512.repulsive.eps}
\end{figure}

\begin{figure}[tbp]
\includegraphics[width=0.45\textwidth]{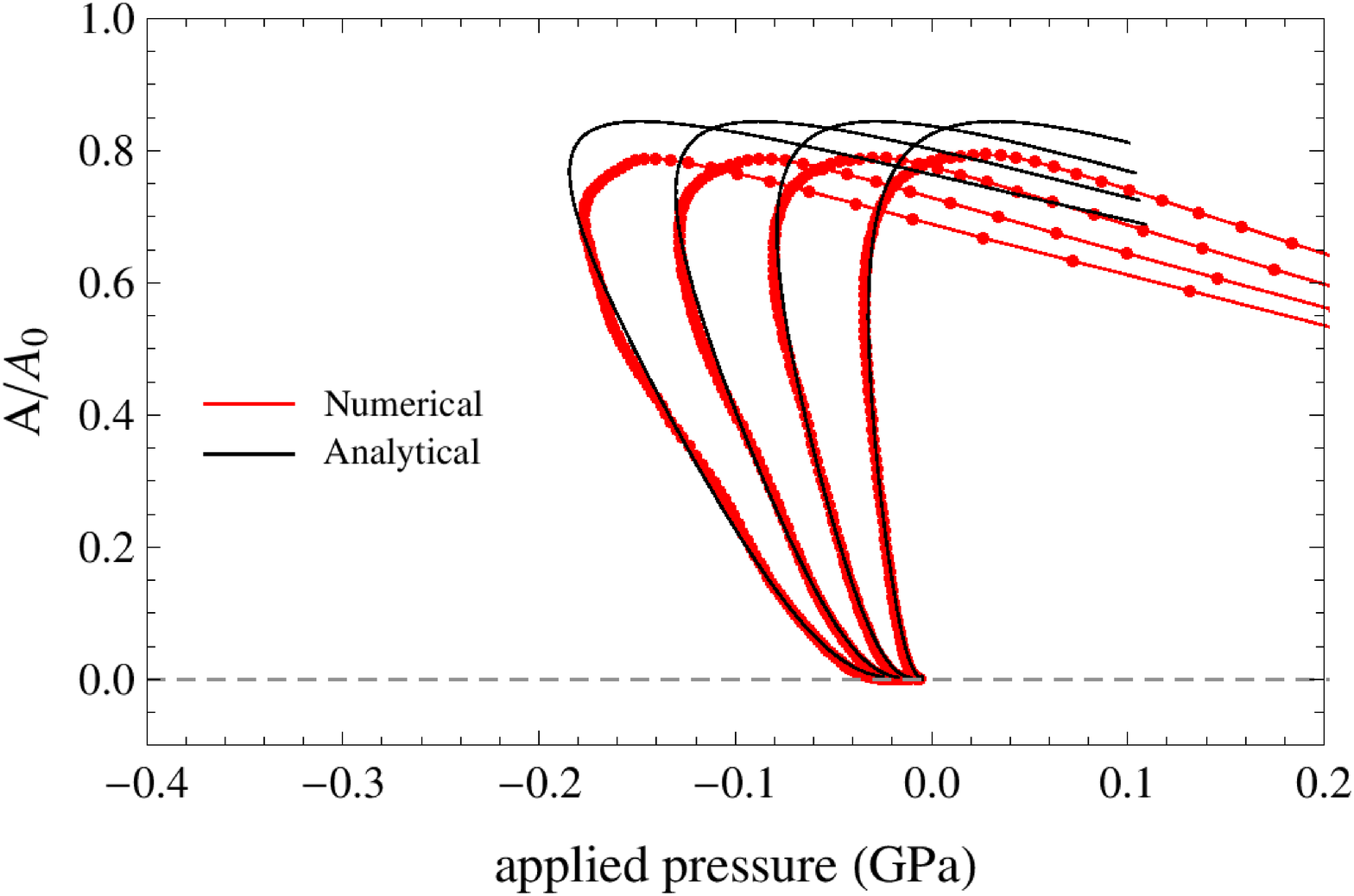}
\caption{Normalized (projected) area of attractive contact $A_{\mathrm{a}}/A_{0}$ as a
function of the applied pressure $p_{\mathrm{N}}$, for different values of
work of adhesion. For the same parameters as in Fig. \protect\ref%
{512.repulsive.eps}. }
\label{512.attractive.eps}
\end{figure}

\begin{figure}[tbp]
\includegraphics[width=0.45\textwidth]{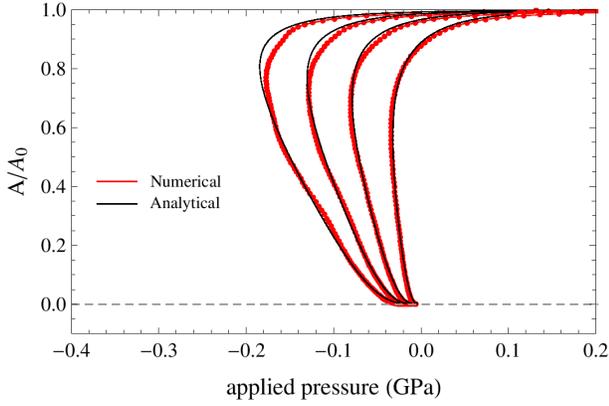}
\caption{Normalized (projected) contact area $A/A_{0}=\left( A_{\mathrm{r}}+A_{\mathrm{a}%
}\right) $ as a function of the applied pressure $p_{\mathrm{N}}$, for
different values of work of adhesion. For the same parameters as in Fig. 
\protect\ref{512.repulsive.eps}. }
\label{512.totalarea.eps}
\end{figure}

In Fig. \ref{512.repulsive.eps}-\ref{512.totalarea.eps} we show,
respectively, the normalized and projected area of repulsive contact $A_{%
\mathrm{r}}/A_{0}$, of attractive contact $A_{\mathrm{a}}/A_{0}$ and the
total interaction area $A/A_{0}=\left( A_{\mathrm{r}}+A_{\mathrm{a}}\right)
/A_{0}$ as a function of the applied (nominal) pressure $p_{\rm N}$. Red dots are
from the deterministic (numerical) model, whereas black solid lines are from the mean field
theory. We note that whilst the repulsive interaction area is slightly
underestimated by the theory, the pull-off pressures are remarkably
accurately captured at the different adopted values of work of adhesion.
Moreover, the total interaction area (see Fig. \ref{512.totalarea.eps}), as
a function of applied pressure, seems to be only marginally affected by the
exact contact boundary conditions adopted in the mean field theory,
resulting in a perfect match with the numerical predictions, as it could
have been expected. It is indeed well known that Persson's contact mechanics
accurately predicts the distribution of interfacial separations\cite{Carlos}.
Hence the total interaction area, which is evaluated from the distribution
of interfacial separation, it is accurately captured too. Also note that the
simulated contact is close to the DMT-limit, as shown in Fig. \ref%
{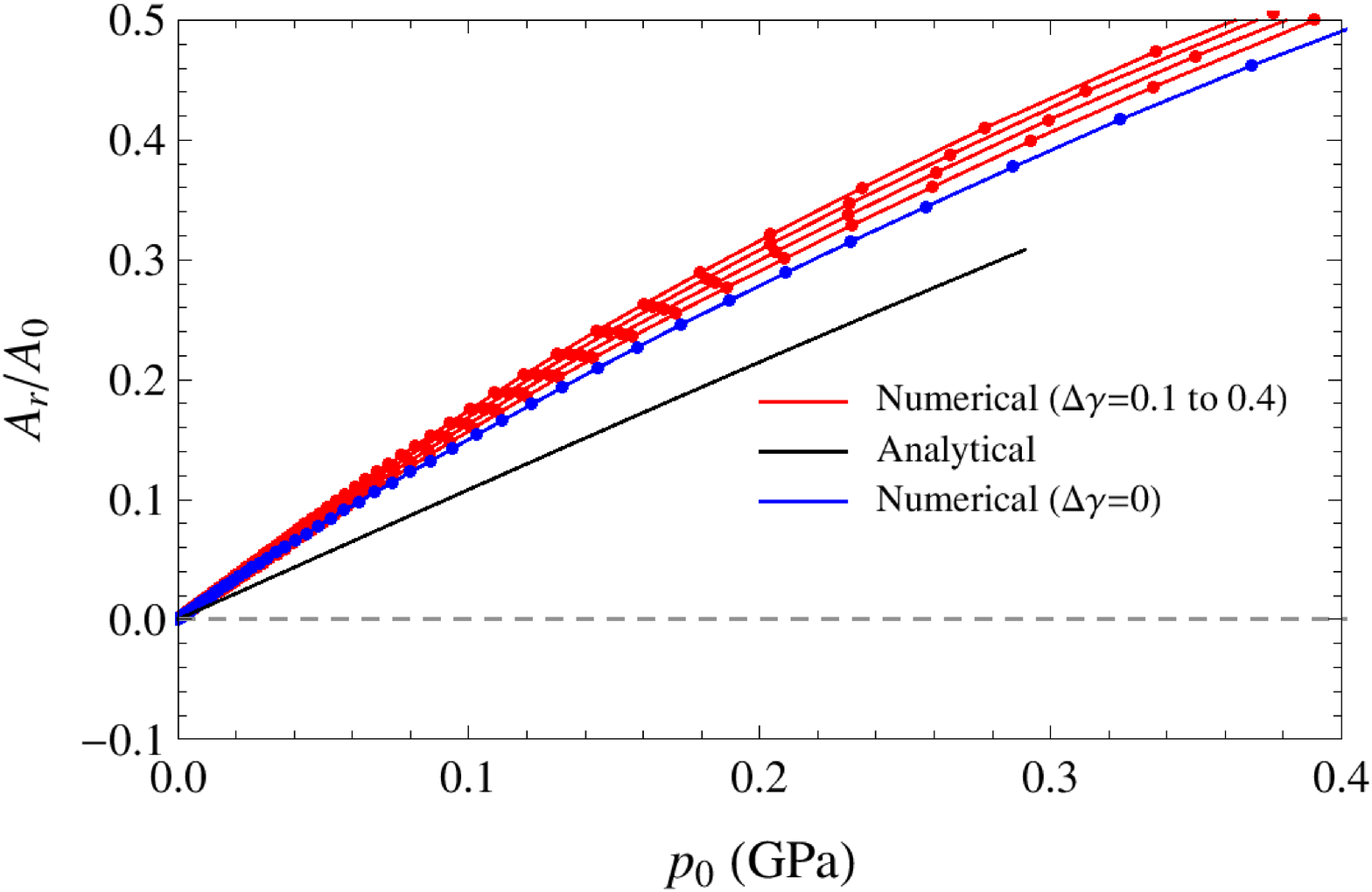}, where the repulsive area is reported as a function of the
nominal repulsive pressure ($p_{0}=p_{\mathrm{N}}+p_{\mathrm{ad}}$).

\begin{figure}[tbp]
\includegraphics[width=0.45\textwidth]{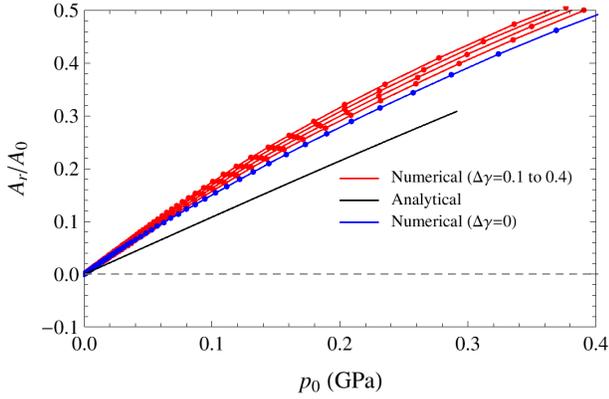}
\caption{Normalized (projected) repulsive area $A_{\mathrm{r}}/A_{0}=$ as a function of
the nominal repulsive pressure $p_{0}=p_{\mathrm{N}}+p_{\mathrm{\mathrm{ad}}%
} $, for different values of work of adhesion. For the same parameters as in
Fig. \protect\ref{512.repulsive.eps}. }
\label{512.DMT.eps}
\end{figure}

\begin{figure}[tbp]
\includegraphics[width=0.45\textwidth]{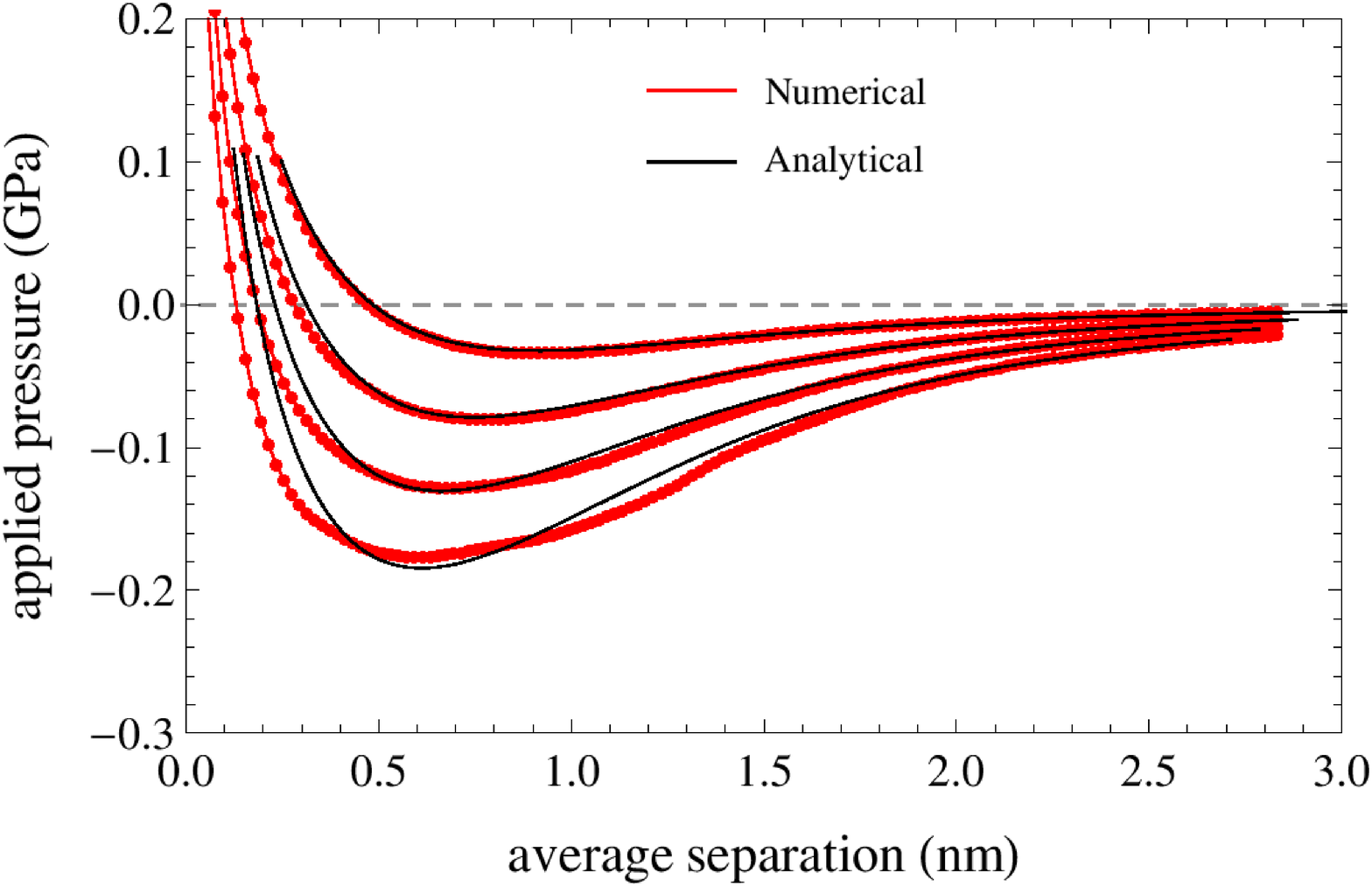}
\caption{Nominal pressure $p_{\text{N}}$ as a function of the average
interfacial separation $\bar{u}$. For the same parameters as in Fig. \protect
\ref{512.repulsive.eps}.}
\label{512.p0.avgsep.eps}
\end{figure}

In Fig. \ref{512.p0.avgsep.eps} we show the applied pressure $p_{\rm N}$ as a function of the average interfacial
separation $\bar{u}$, for different values of $\Delta \gamma $, as determined from the theory (black
curves) and the numerical model. As expected from the previous arguments, the
agreement is remarkably good in almost the entire range of average
interfacial separations.

The power spectral density can be nowadays routinely obtained with commonly
available lab profilometers. However, usually one has to adopt different
acquisition techniques depending on the range of roughness length scales
needed to be investigated. Therefore, it would be particularly interesting
to appreciate the extent to which the macroscopic adhesive characteristics,
such as pull-off pressure, depends on the effect of adding (or, inversely,
not measuring) an increasing number of surface roughness frequency
components. To do so, we gradually extend the numerically calculated
roughness spectral components of Fig. \ref{psd.eps}, as shown in Fig. \ref%
{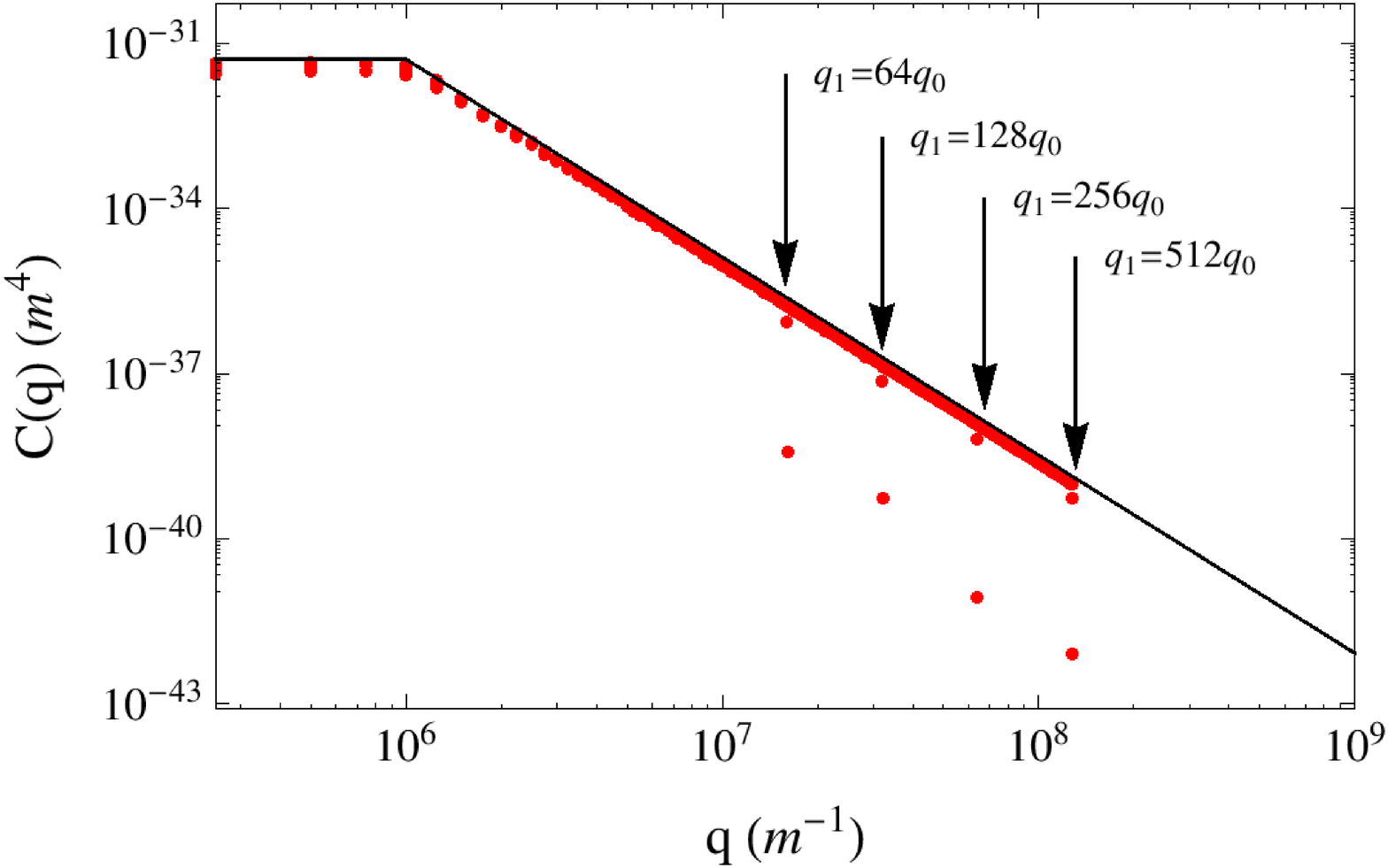}, up to a system size of $2^{24}$ mesh points. 
In Fig. \ref{nx.repulsive.eps}-\ref{nx.totalarea.eps} we show, respectively, the
normalized and projected area of repulsive contact $A_{\mathrm{r}}/A_{0}$,
the attractive contact $A_{\mathrm{a}}/A_{0}$ and the total interaction area 
$A/A_{0}=\left( A_{\mathrm{r}}+A_{\mathrm{a}}\right) /A_{0}$ as a function
of the applied nominal pressure $p_{\rm N}$, for different truncation
wavevectors. Red dots are the predictions of the numerical model, whereas
black solid lines are from the mean field theory. The pull-off pressure is
almost independent of the large-wavevector content of the PSD, whereas the
repulsive contact area, as expected, decreases by including large-wavevector
(small wavelength) roughness. Moreover, the large-wavevector roughness does
not contribute significantly to the $h_{\text{\textrm{rms}}}$, as is clear
both theoretically and numerically from Fig. \ref{nx.p0.avgsep.eps}, where
the applied pressure is reported as a function of the average interfacial
separation.

Finally, let us compare the theory prediction with numerical results for the
effective interfacial energy $\gamma_{\mathrm{eff}}$. In Fig. \ref%
{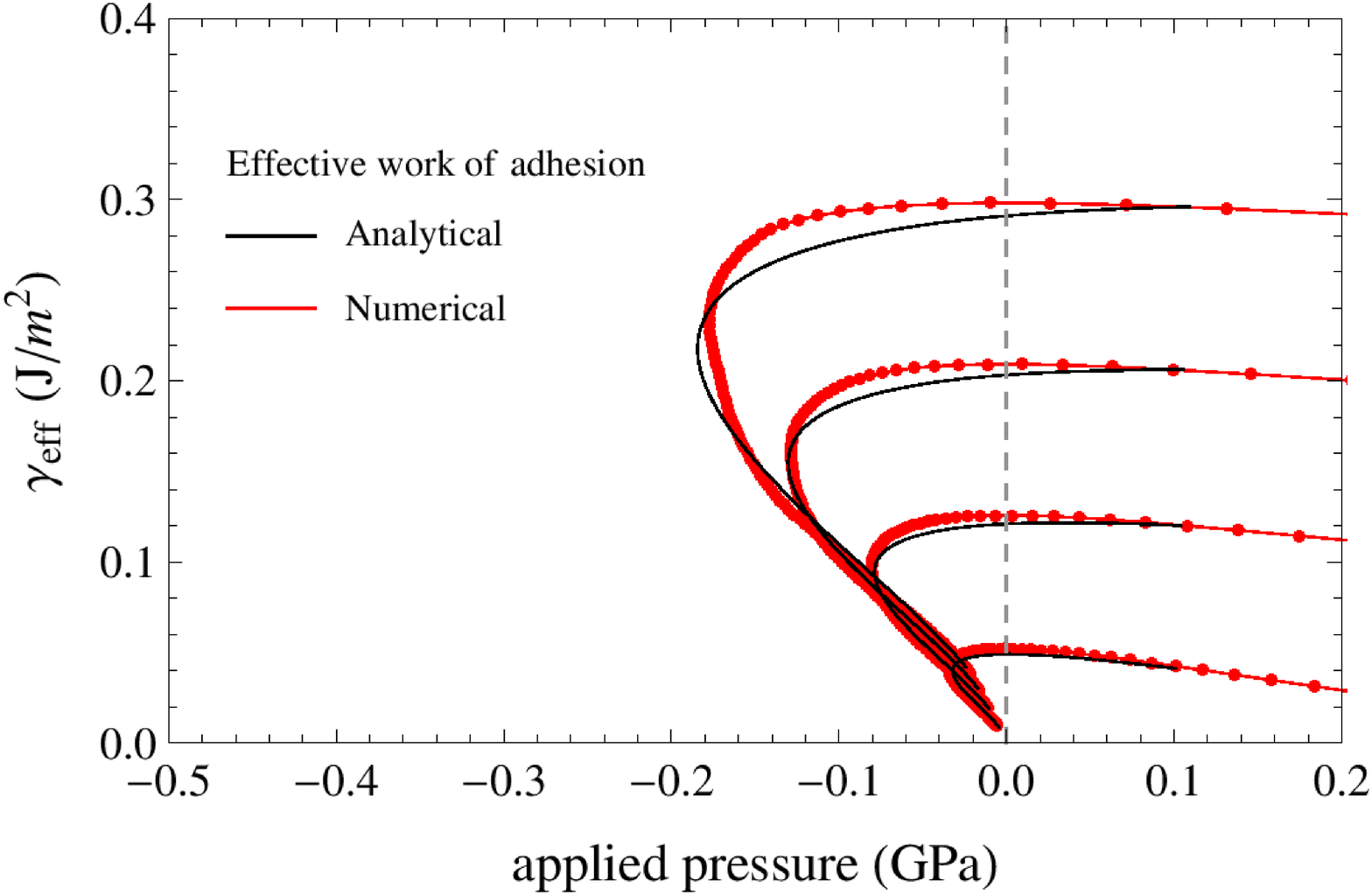} the effective interfacial energy $\gamma_{\mathrm{%
eff}} = [E_{\mathrm{ad}}-U_{\mathrm{el}}]/A_0$ [where $E_{\mathrm{ad}}$ is
the (attractive) Van der Waals interaction energy and $U_{\mathrm{el}}$ the
(repulsive) elastic deformation energy] as a function of the nominal
pressure $p_{\mathrm{N}}$ acting on the block. Results are shown for the
work of adhesion $\Delta \gamma = 0.1$, $0.2$, $0.3$ and $0.4 \ \mathrm{J/m^2%
}$. The red data points are from the exact numerical simulation and the black
lines from the DMT-like theory (Sec. 3.2) also shown in Fig. \ref%
{combine.p.A.p.gamma.eps}. The elastic solid Young's modulus $E=10^{12} \ 
\mathrm{Pa}$ and Poisson number $\nu = 0.5$.

\begin{figure}[tbp]
\includegraphics[width=0.45\textwidth]{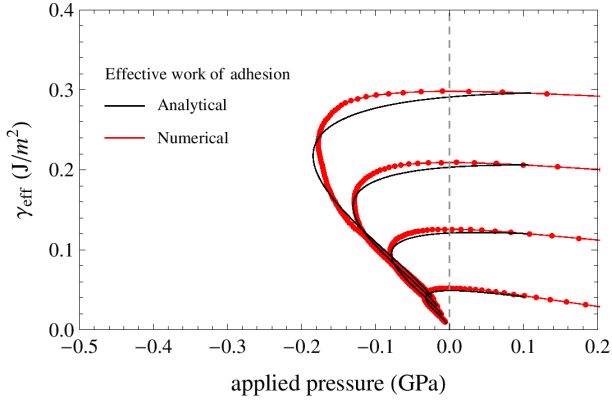}
\caption{The effective interfacial energy $\protect\gamma _{\mathrm{eff}%
}=[E_{\mathrm{ad}}-U_{\mathrm{el}}]/A_{0}$ [where $E_{\mathrm{ad}}$ is the
(attractive) Van der Waals interaction energy and $U_{\mathrm{el}}$ the
(repulsive) elastic deformation energy] as a function of the applied
pressure $p_{\mathrm{N}}$ acting on the block. Results are shown for the
work of adhesion $\Delta \protect\gamma =0.1$, $0.2$, $0.3$ and $0.4\ 
\mathrm{J/m^{2}}$. The red data points are from the exact numerical
simulation and the black lines from the DMT-like theory (Sec. 3.2) also
shown in Fig. \protect\ref{combine.p.A.p.gamma.eps}. For the same parameters as in Fig. \protect
\ref{512.repulsive.eps}. }
\label{512.workofadhesion.eps}
\end{figure}

In Fig. \ref{nx.workofadhesion.eps} we show similar results for the
effective interfacial energy $\gamma _{\mathrm{eff}}$ but now for $\Delta
\gamma =0.2\ \mathrm{J/m^{2}}$, and for several large wavevector cut-off $%
q_{1}=64q_{0}$, $128q_{0}$ and $256q_{0}$. Note that the effective
interfacial energy $\gamma _{\mathrm{eff}}$ is rather insensitive to the
large wavevector cut-off $q_{1}$. The reason for this is that the repulsive
elastic energy $U_{\mathrm{el}}$ is dominated by the long-wavelength
roughness. In both figures \ref{512.workofadhesion.eps} and \ref%
{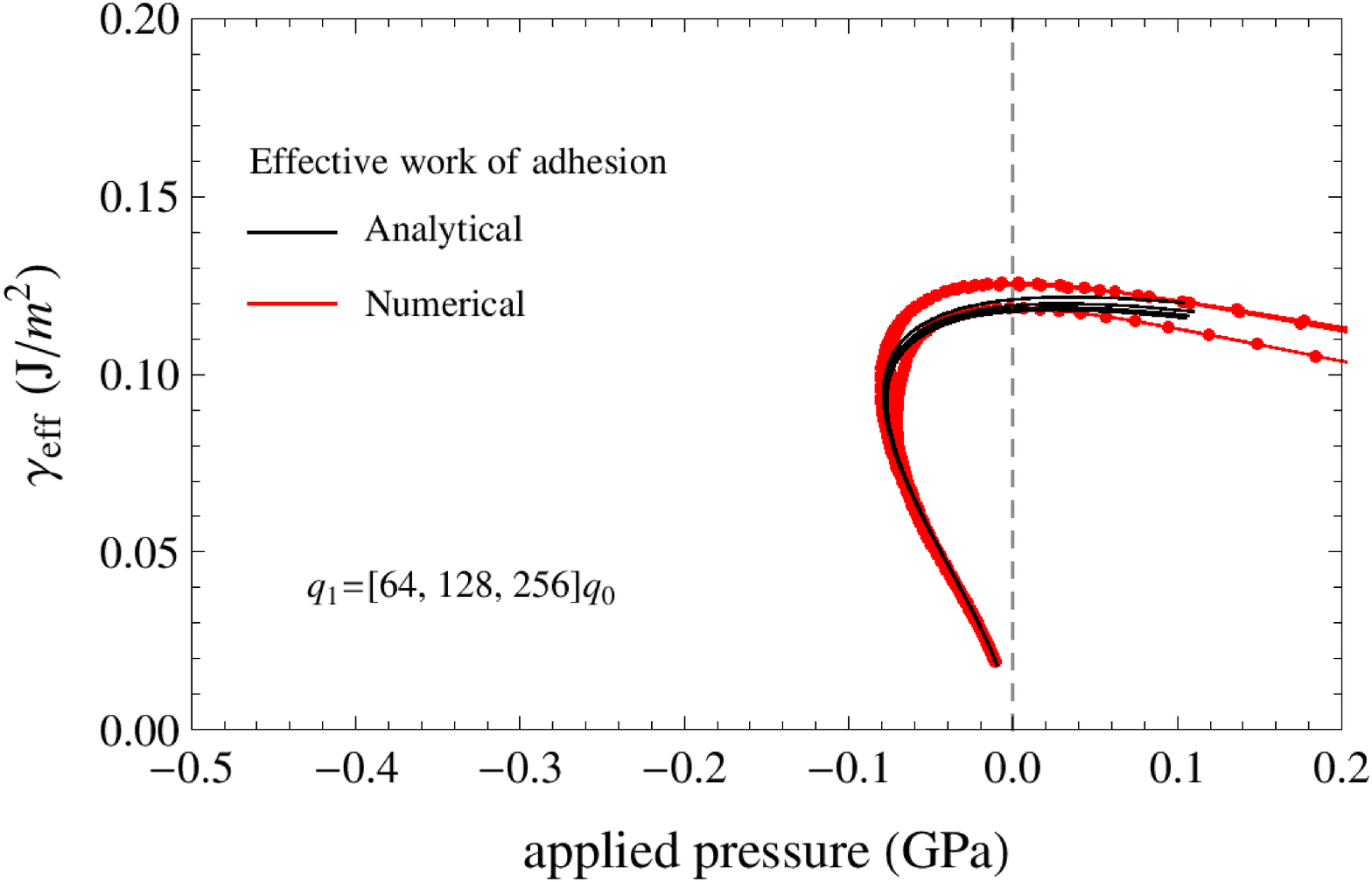} there is remarkable good agreement between theory
and the simulations.

\begin{figure}[tbp]
\includegraphics[width=0.45\textwidth]{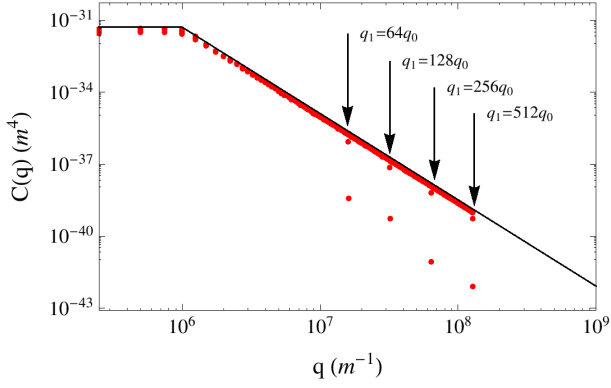}
\caption{Power spectral density $C\left( q\right) $ as a function of $q$
(solid black line). For an isotropic surface roughness with cut-off $%
q_{0}=q_{\mathrm{r}}/4$ and root-mean-square roughness $h_{\mathrm{rms}%
}=0.6\ \mathrm{nm}$, and with self-affine regime in the frequency range $q_{%
\mathrm{r}}=10^{6}\mathrm{m}^{-1}$ to $q_{1}=10^{3}q_{\mathrm{r}}$ ($H=0.8$%
). The numerical adopted PSD (red dots) is truncated at $q_{1}=[64,\ 128,\
256,\ 512]q_{0}$, with 8 divisions at the smallest lengh scale ($q_{1}$). }
\label{nx.psd.eps}
\end{figure}

\begin{figure}[tbp]
\includegraphics[width=0.45\textwidth]{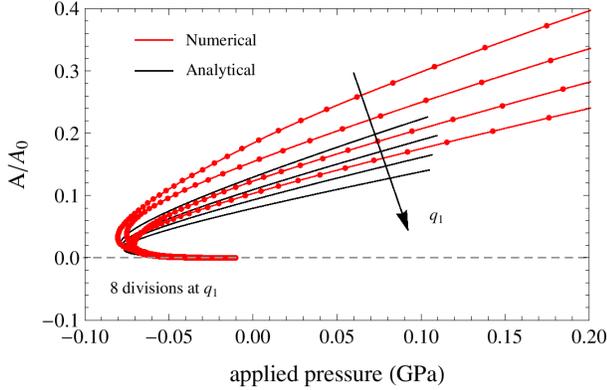}
\caption{Normalized (projected) area of repulsive contact $A_{\mathrm{r}}/A_{0}$ as a
function of the applied pressure $p_{\mathrm{N}}$, and for $\Delta \protect%
\gamma =0.2\ \mathrm{J/m^{2}}$. For an elastic solid with $E_{\mathrm{r}%
}=1.33\times 10^{12}\mathrm{GPa}$ and for the surface roughness of Fig. \protect
\ref{nx.psd.eps} (with $q_{0}=2.5\ 10^{5}\mathrm{m}^{-1}$, $q_{\mathrm{r}%
}=4q_{0}$, $H=0.8$, resulting in $C_{0}=5.24~10^{-32}\mathrm{m}^{4}$), at
different truncation frequencies $q_{1}=64,128,256,512\ q_{0}$.}
\label{nx.repulsive.eps}
\end{figure}

\begin{figure}[tbp]
\includegraphics[width=0.45\textwidth]{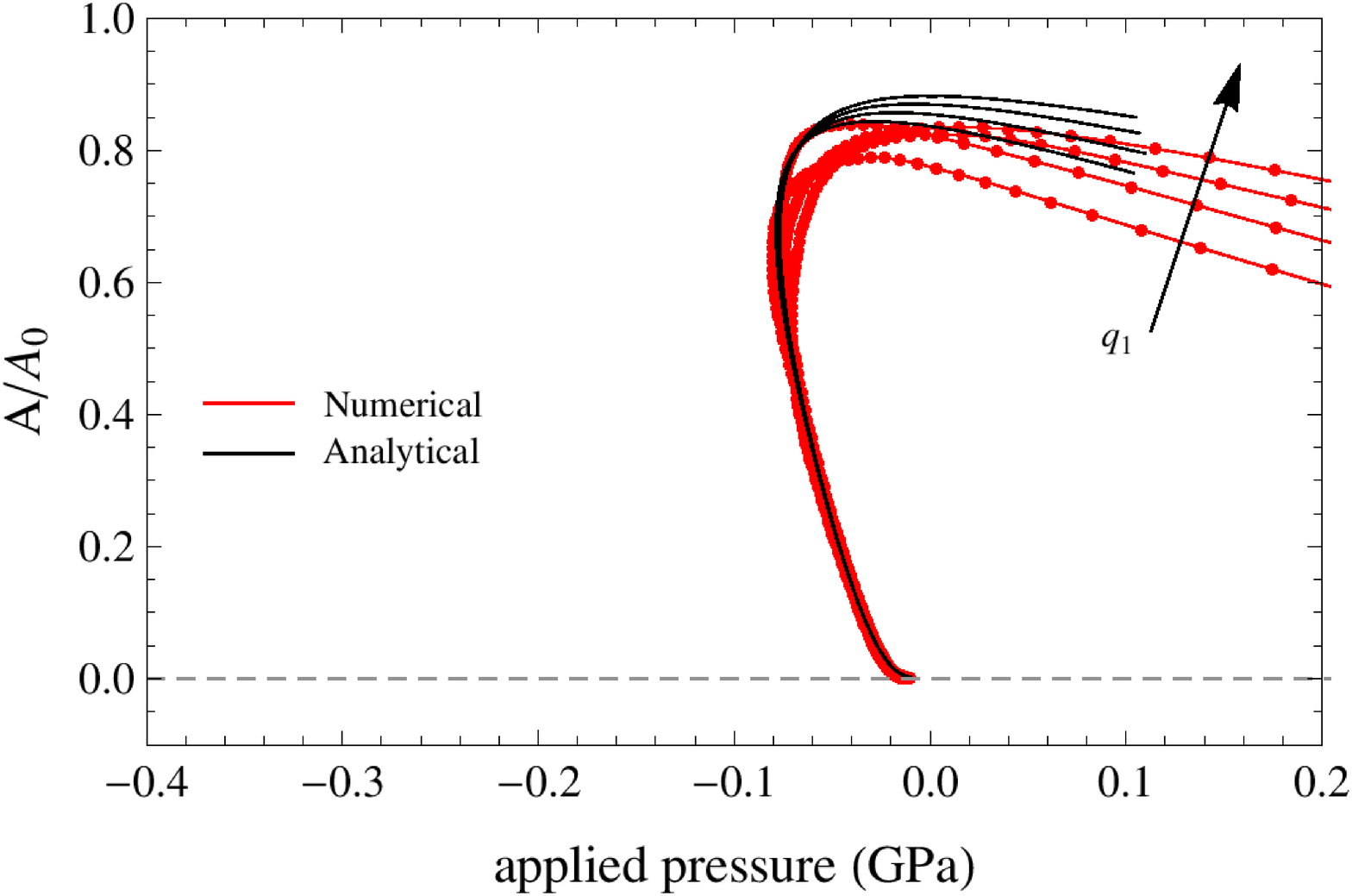}
\caption{Normalized (projected) area of attractive contact $A_{\mathrm{\mathrm{a}}%
}/A_{0} $ as a function of the applied pressure $p_{\mathrm{N}}$. For the
same parameters as in Fig. \protect\ref{nx.repulsive.eps}.}
\label{nx.attractive.eps}
\end{figure}

\begin{figure}[tbp]
\includegraphics[width=0.45\textwidth]{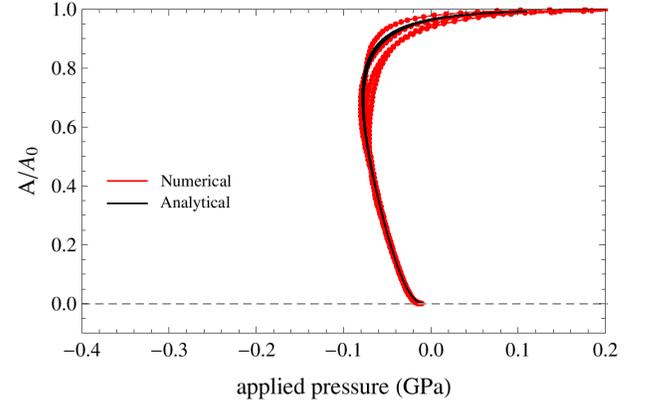}
\caption{Normalized (projected) contact area $A/A_{0}=\left( A_{\mathrm{r}}+A_{\mathrm{a}%
}\right) $ as a function of the applied pressure $p_{\mathrm{N}}$. For the
same parameters as in Fig. \protect\ref{nx.repulsive.eps}. }
\label{nx.totalarea.eps}
\end{figure}

\begin{figure}[tbp]
\includegraphics[width=0.45\textwidth]{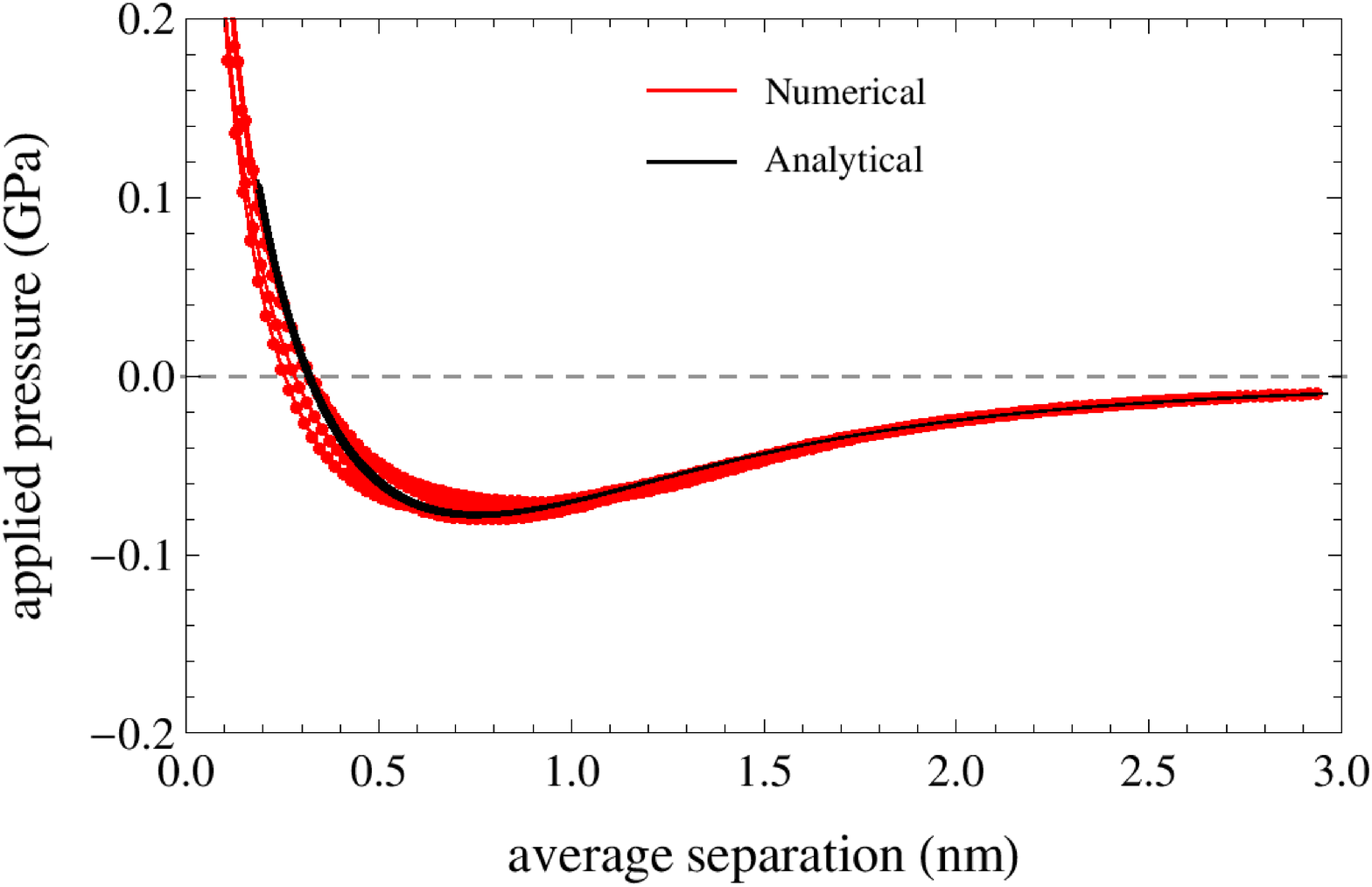}
\caption{Applied pressure $p_{\mathrm{N}}$ as a function of the average
interfacial separation $\bar{u}$. For the same parameters as in Fig. \protect
\ref{nx.repulsive.eps}. }
\label{nx.p0.avgsep.eps}
\end{figure}

\begin{figure}[tbp]
\includegraphics[width=0.45\textwidth]{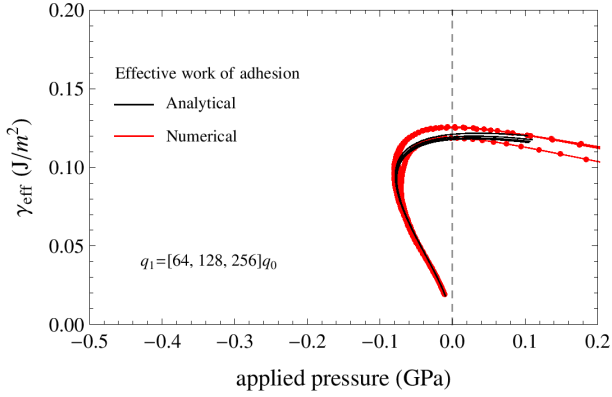}
\caption{The effective interfacial energy $\protect\gamma _{\mathrm{eff}%
}=[E_{\mathrm{ad}}-U_{\mathrm{el}}]/A_{0}$ [where $E_{\mathrm{ad}}$ is the
(attractive) Van der Waals interaction energy and $U_{\mathrm{el}}$ the
(repulsive) elastic deformation energy] as a function of the applied
pressure $p_{\mathrm{N}}$ acting on the block. Results are shown for the
large wavevector cut-off $q_{1}=64q_{0}$, $128q_{0}$ and $256q_{0}$. For the
work of adhesion $\Delta \protect\gamma =0.2\ \mathrm{J/m^{2}}$. The red
data points are from the exact numerical simulation and the black lines from
the DMT-like theory (Sec. 3.2). For the same parameters as in Fig. \protect
\ref{nx.repulsive.eps}. }
\label{nx.workofadhesion.eps}
\end{figure}

\vskip0.3cm \textbf{6 Discussion}

In the discussion above we have neglected adhesion hysteresis. Adhesion
hysteresis is particular important for viscoelastic solids such as most
rubber compounds. However, even for elastic solids adhesion hysteresis may
occur. Thus, not all the stored elastic energy $U_{\mathrm{el}}$ may be used
to break adhesive bonds during pull-off but some fraction of it may be
radiated as elastic waves (phonons) into the solids. This would result in an
increase in the effective interfacial binding energy during pull-off, and
would result in adhesion hysteresis.

We note that adhesion hysteresis is observed already for smooth surfaces in
the JKR-limit (elastically soft solids) but not in the DMT-limit (hard
solids)\cite{comment}. Since for randomly rough surfaces the contact
mechanics may be close to the DMT-limit for short length scales (high
resolution) while close to the JKR-limit at large enough length scales, as
in Fig. \ref{1logq.2log.Tabor.Length.eps}, one expects in many cases that the
bond-breaking process involved at short length scale is reversible (no
hysteresis), while the elastic deformations at large enough length scales
show hysteresis, involving rapid (dissipative) processes during pull-off.

Contact mechanics for randomly rough surfaces is a hard problem to treat
numerically in the JKR-limit (see Appendix A) and most
studies published are close to the DMT-limit. While this case may be
relevant for many hard materials, most adhesion experiments involves soft
materials like silicon rubber (PDMS). In this case the adhesion will be
JKR-like in a large range of length scales.

We note that adhesion problems which are JKR-like for large length scales
and DMT-like for short length scales can be approximately treated using the
theory presented above: We plot the Tabor length $d_{\mathrm{T}}(q)$ as a
function of $\mathrm{log} q$ as in Fig. \ref{1logq.2log.Tabor.Length.eps} and
divide the $\mathrm{log} q$ axis into a large wavevector region $q>q^*$ and
a short wavevector region $q<q^*$ where $d_{\mathrm{T}}(q^*) = d_{\mathrm{c}%
} $. We use the DMT-like theory to calculate $\gamma_{\mathrm{eff}}(q^*)$
including only the roughness components with $q>q^*$. Next we apply the
JKR-like theory for the $q<q^*$ region with $\Delta \gamma = \gamma_{\mathrm{%
eff}} (q^*)$. This treatment is of course only approximate since there will
be a region close to $q=q^*$ which is neither DMT-like nor JKR-like, but if
this region (on the $\mathrm{log} q$-scale) is small compared to the total
decades of length scales involved it may constitute a good approximation.
This picture of adhesion is similar to the Renormalization Group (RG)
procedure used in statistical physics where short wavelength degrees of
freedom (here the short wavelength roughness involved in the DMT-like
contact mechanics) are integrated out (removed) to obtain effective
equations relevant at the macroscopic length scale (here the JKR-like
contact mechanics). When applying the RG procedure one often finds that
processes or phenomena which appear very different at the microscopic (say
atomistic) limit result in the same macroscopic equations of motion e.g.,
the Navier Stokes equations of fluid flow does not really depend on the
exact nature of the force law between the atoms or molecules except it
determines or influence the fluid density and viscosity. Similar, for large
surface roughness the force law between the surfaces, which is important at
short length scale (DMT-limit) does not really matter for the macroscopic
(JKR-like) contact mechanics except it determines the effective interfacial
binding energy $\Delta \gamma = \gamma_{\mathrm{eff}}(q^*)$ to be used in
the JKR theory. This statement does not hold when the surface roughness
amplitude is very small, such as in the present study, because the (average)
surface separation in the non-contact area is only of order $\sim 1 \ 
\mathrm{nm}$ and at this separation the wall-wall interaction potential is
still important, in particular for small index $n$. For charged bodies, due
to the long-range of the coulomb interaction, the wall-wall interaction potential is
important for any wall-wall separation.

\vskip 0.3cm \textbf{7 Summary}

We have discussed how surface roughness influence the adhesion between elastic
solids. We have introduced a Tabor number which depends on the length scale
or magnification, and which gives information about the nature of the
adhesion at different length scales. In most cases the contact mechanics
will be DMT-like at short length scales and JKR-like at large length scales.
We have considered two limiting cases relevant for (a) elastically hard
solids with weak adhesive interaction (DMT-limit) and (b) elastically soft
solids or strong adhesive interaction (JKR-limit). For the former cases we
have studied the nature of the adhesion using different adhesive force laws (%
$F\sim u^{-n}$, $n=1.5-4$, where $u$ is the wall-wall separation) and by
comparing the mean field theory predictions with the results of exact
numerical calculations. The theory results have been compared to the results
of exact numerical simulations, and good agreement between theory and the
simulation results was obtained.

\vskip 0.5cm \textbf{Appendix A: Numerical model} 
\label{appendix.1} \setcounter{equation}{0}

We consider the case of two elastic solids patterned with random or
deterministic roughness. We assume the generic roughness to be characterized
by a small wavelength cut-off $q_{0}=2\pi /L_{0}$ with $L_{0}\ll L$, where $%
L $ is the representative size of the macroscopic contact region between the
two solids. Given such a large difference of length scales, we can easily
identify a representative elementary volume (RVE) of interface of length
scale $L_{\mathrm{RVE}}$, with $L_{0}\ll L_{\mathrm{RVE}}\ll L$, over which
we can average out the contact mechanics occurring at smaller length scales
(say, at $\lambda \ll L_{\mathrm{RVE}}$). Note that the numerical or
analytical homogenization of the high-frequency content of a generic
physical medium/process model is very common in physics and engineering,
since it allows to build a mean field formulation of the model itself,
characterized by effective (i.e. smoother) physical properties, varying over
length scales of order $\sim L_{0}$. This is e.g. the case of the rough
contact mechanics, where the accurate knowledge of the relationship between
the effective interfacial characteristics (average interfacial separation,
effective work of adhesion, etc., to cite few), plays a fundamental role in
many physical processes, from friction and thermal/electrical conduction, to
adhesion and interfacial fluid flow. Here we briefly describe the novel efficient
numerical approach devoted to simulate the contact mechanics of
realistically-rough interfaces at the REV scale.

\begin{figure}[tbp]
\includegraphics[width=0.45\textwidth]{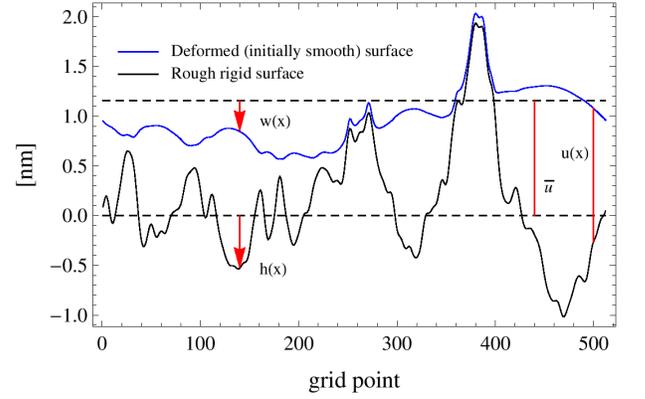}
\caption{{}Description of the gap [see Eq. (\protect\ref{A1.gap})] resulting
from a generic cross section of the contact interface.}
\label{contact.geometry.eps}
\end{figure}

In Fig. \ref{contact.geometry.eps} we show a schematic of the contact geometry.
We assume the contact to occur under isothermal conditions, and the
roughness to be characterized by a small mean square slope, in order to make
use of the well known half space theory. Moreover, the roughness is assumed
to be periodic with period $L_{0}$ in both $x$- and $y$-direction. The local
separation between the mating interfaces $u\left( \mathbf{x}\right) $ is
shown in Fig. \ref{contact.geometry.eps}, and it can be immediately agreed to be:%
\begin{equation}
u\left( \mathbf{x}\right) =\bar{u}+w\left( \mathbf{x}\right) -h\left( 
\mathbf{x}\right) ,\MSlabel{A1.gap}
\end{equation}%
where $\bar{u}$ is the average interfacial separation, $w\left( \mathbf{x}%
\right) $ the surface out-of-average-plane displacement and $h\left( \mathbf{%
x}\right) $ the surface roughness, with $\left\langle w\left( \mathbf{x}%
\right) \right\rangle =\left\langle h\left( \mathbf{x}\right) \right\rangle
=0$. By defining 
$$w\left( \mathbf{q}\right) =\left( 2\pi \right) ^{-2}\int
d^{2}\mathbf{x\ }w\left( \mathbf{x}\right) e^{-i\mathbf{q}\cdot \mathbf{x}}$$
and 
$$\sigma \left( \mathbf{q}\right) =\left( 2\pi \right) ^{-2}\int
d^{2}\mathbf{x\ }\sigma \left( \mathbf{x}\right)%
e^{-i\mathbf{q}\cdot \mathbf{x}},$$ 
where $\sigma \left( \mathbf{x}%
\right) $ is the distribution of interfacial pressures, it is
(relatively) easy to show that $w\left( \mathbf{x}\right) $ can be related
to $\sigma \left( \mathbf{x}\right) $ through a very simple equation in the
Fourier space:%
\begin{equation*}
w\left( \mathbf{q}\right) =M_{\mathrm{zz}}\left( \mathbf{q}\right) 
\sigma \left( \mathbf{q}\right) ,
\end{equation*}%
where $M_{\mathrm{zz}}\left( \mathbf{q}\right) =-2/\left( \left\vert \mathbf{%
q}\right\vert E_{r}\right) $ for the elastic half space [$M_{\mathrm{zz}%
}\left( \mathbf{q}\right) $ can be equally determined for layered or
viscoelastic materials, for which the reader is referred to Ref. \cite{layer}%
]. Finally, the relation between separation $u\left( \mathbf{x}\right) $ and
interaction pressure $\sigma \left( \mathbf{x}\right) $ is calculated within
the Derjaguin's approximation\cite{guin}, and it can be written in term of a
generic interaction law $\sigma \left( u\right) =f\left( u\right) $. $%
f\left( u\right) $ will be repulsive for $u\leq u_{\mathrm{w}}$ and
attractive otherwise, where $u_{\mathrm{w}}$ is a separation threshold
describing the ideal equilibrium separation. In this work we have adopted
the L-J potential to describe the attractive interaction, but one can
equally make use of different interaction laws (e.g. the Morse potential,
for chemical bonds). The attractive side of $f\left( u\right) $, $f_{\mathrm{%
a}}\left( u\right) $, reads:%
\begin{align}
f_{\mathrm{a}}\left( u\right) & =\frac{8}{3}\frac{\Delta \gamma }{d_{\mathrm{%
c}}}\left[ \varepsilon ^{-9}-\varepsilon ^{-3}\right] \MSlabel{A1.attractive} \\
\varepsilon \left( u\right) & =\left( u-u_{\mathrm{w}}+d_{\mathrm{c}}\right)
/d_{\mathrm{c}},  \notag
\end{align}%
whereas the repulsive $f_{\mathrm{r}}\left( u\right) $:%
\begin{align}
f_{\mathrm{r}}\left( u\right) & =\frac{8}{3}\frac{\Delta \gamma }{d_{\mathrm{%
c}}}\left[ \varepsilon ^{-9}-\varepsilon ^{-3}\right] \MSlabel{A1.repulsive} \\
\varepsilon \left( u\right) & =u/u_{\mathrm{w}}.  \notag
\end{align}%
Usually we adopt $u_{\mathrm{w}}\ll d_{\mathrm{c}}$. Note however that for $%
u_{\mathrm{w}}\rightarrow 0$ the repulsive term converges to an hard wall,
whereas for $u_{\mathrm{w}}=d_{\mathrm{c}}$ we return to the classical
(integrated)\ L-J interaction law.

\begin{figure}[tbp]
\includegraphics[width=0.45\textwidth]{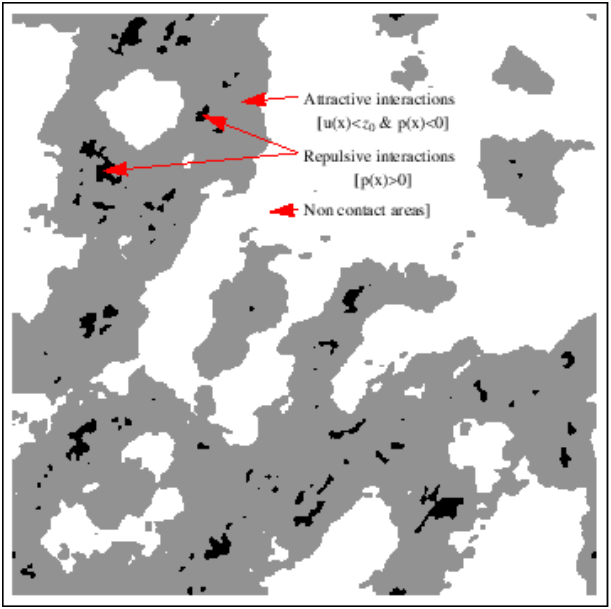}
\caption{{}Example of contact map.}
\label{contact.area.eps}
\end{figure}

Eqs. (\ref{A1.gap}), (\ref{A1.attractive}) and (\ref{A1.repulsive}) are
discretized on a regular square mesh of grid size $\delta $, resulting in
the following set of equations:%
\begin{align}
L_{ij}& =-u_{ij}+\left( \bar{u}+w_{ij}-h_{ij}\right) \MSlabel{A1.Lij} \\
\sigma _{ij}& =f_{\mathrm{a}}\left( u_{ij}\right) +f_{\mathrm{r}}\left(
u_{ij}\right) \MSlabel{A1.f(uij)} \\
\sigma \left( x_{ij}\right) & \rightarrow \Delta \sigma \left( q_{hk}\right)
=M_{\mathrm{zz}}^{-1}\left( q_{hk}\right) w\left( q_{hk}\right) \rightarrow
w\left( x_{ij}\right) ,\MSlabel{A1.integration}
\end{align}%
where $L_{ij}$ is the generic residual (related to the generic iterative
solution $u_{ij}$). In order to solve Eqs. (\ref{A1.Lij})-(\ref{A1.integration})%
, we rephrase Eq. \ref{A1.Lij} in term of the following ideal Molecular
Dynamics process%
\begin{equation}
\ddot{u}_{ij}+2\xi _{ij}\omega _{ij}\dot{u}_{ij}=\omega _{ij}^{2}L_{ij}%
\MSlabel{A1.MD}
\end{equation}%
we solve with a velocity Verlet integration scheme. $\xi _{ij}$ and $\omega
_{ij}$ are, respectively, the generic damping factor and modal frequency of
the Residuals Molecular Dynamics system (RMD), which can be smartly used to
damp the error dynamics. Therefore, at equilibrium ($\ddot{u}_{ij}=\dot{u}%
_{ij}=0$), Eq. (\ref{A1.MD}) returns the solution of Eqs. (\ref{A1.Lij})-(\ref%
{A1.integration}) at zero residuals. The adoption of the RMD scheme allows
for the (ideal time) search of the solution to move in a (generic)
non-physical error space to finally furnish, at equilibrium, the targeted
(zero residuals) solution. We have found that this very efficiently avoids
to be trapped in slow relaxation dynamics and/or non-physical (uncorvergent)
solution as otherwise obtained with classical (usually very slow) relaxation
approaches (e.g. under-relaxation, often adopted in the literature for
smooth contact conditions, see e.g. Ref. \cite{congrui}) applied to
realistically-rough interacting surfaces. The solution accuracy is set by
requiring%
\begin{align}
\left\langle L_{ij}^{2}/u_{ij}^{2}\right\rangle ^{1/2}& <\varepsilon _{%
\mathrm{L}}\MSlabel{A1.err} \\
\left\langle \left[ \left( u_{ij}^{n}-u_{ij}^{n-1}\right) /u_{ij}^{n-1}%
\right] ^{2}\right\rangle ^{1/2}& <\varepsilon _{\mathrm{u}},  \notag
\end{align}%
where both errors are typically of order $10^{-4}$. As anticipated in
previous sections, the nominal projected contact area 
$A/A_{0}$ is given by $A_{\mathrm{r}}/A_{0}+A_{\mathrm{a}}/A_{0}$, where $A_{%
\mathrm{r}}$ is the area of repulsive interaction (defined by\ $\sigma
\left( \mathbf{x}\right) >0$), and where $A_{\mathrm{a}}$ is the area of
attractive interaction (defined by $u\left( x\right) -u_{\mathrm{w}}<d_{%
\mathrm{c}}$ and $\sigma \left( \mathbf{x}\right) <0$). In Fig. \ref%
{contact.area.eps} we show a typical contact area map for a DMT rough
interaction, where $A_{\mathrm{r}}/A_{0}$ and $A_{\mathrm{a}}/A_{0}$
correspond, respectively, to black and gray domains.

We also observe that for any discretized formulation of the adhesive contact
mechanics, a fracture tensile stress can be related to the mesh size
characteristics of the contact. To determine it, we make use of the
penny-shaped crack solution (see e.g. Ref. \cite{P3}), whose tensile stress $%
\sigma _{\mathrm{a}}$ reads (the $\pi /4$ takes into account the square
shape of the grid):%
\begin{equation*}
\sigma _{\mathrm{a}}=\frac{\pi }{4}\sqrt{\pi \Delta \gamma E_{\mathrm{r}%
}/\delta }.
\end{equation*}%
$\sigma _{\mathrm{a}}$ has to be compared to the maximum tensile stress
given by the interaction law, $\sigma _{\mathrm{t}}=\Delta \gamma /d_{%
\mathrm{c}}$ and, in particular, a detachment parameter $\varepsilon _{%
\mathrm{a}}=\sigma _{\mathrm{a}}/\sigma _{\mathrm{t}}>1$ in order to
guarantee the convergence of the numerical solution. By adopting a Tabor
number definition $\mu _{\mathrm{T,\lambda }}=\left( \frac{\lambda ^{2}}{A_{%
\mathrm{\lambda }}\left( 2\pi \right) ^{2}}{\frac{\Delta \gamma ^{2}}{E_{%
\mathrm{r}}^{2}d_{\mathrm{c}}^{3}}}\right) ^{1/3}${, where }$\lambda $ is
smallest roughness wavelentgh and $A_{\mathrm{\lambda }}$ the corresponding
amplitude, a convergent numerical solution will be achieved if%
\begin{equation}
n_{\mathrm{\lambda }}=\lambda /\delta >\frac{32}{\pi ^{2}}\mu _{\mathrm{%
T,\lambda }}^{3/2}\left( \frac{A_{\mathrm{\lambda }}}{d_{\mathrm{c}}}\right)
^{1/2},\MSlabel{A1.criterion}
\end{equation}%
where $n_{\mathrm{\lambda }}$ is the number of discretization points at
wavelength $\lambda $. It is interesting to observe from Eq. (\ref%
{A1.criterion}) that large Tabor numbers, i.e. adhesive interactions
occurring in the full JKR regime, are numerically harsh to be modelled [due
to the fine mesh description required to satisfy Eq. (\ref{A1.criterion}):
e.g. for $\mu _{\mathrm{T,\lambda }}=10$ and $A_{\mathrm{\lambda }}/d_{%
\mathrm{c}}=10^{2}$, we have $n_{\mathrm{\lambda }}\approx 10^{3}$].
However, as recently shown\cite{Mart}, a JKR regime can be conveniently obtained for 
$\mu _{\mathrm{T}}$ values close to $1$, reducing the computational
complexity of JKR\ interactions.

\begin{figure}[tbp]
\includegraphics[width=0.45\textwidth]{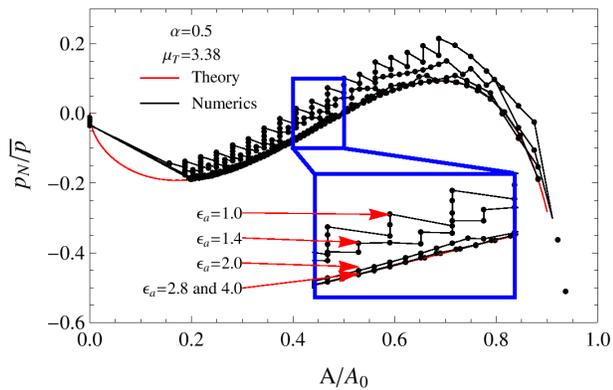}
\caption{{}Dimensionless applied pressure $p_{\mathrm{N}}/\bar{p}$ as a
function of the contact area, for a Westergaard like contact geometry. Red
curve is from Johnson's theory, whereas dots are the corresponding numerical
predictions at different detachment parameters. $\protect\mu _{\mathrm{T,%
\protect\lambda }}=3.4$ for all numerical results.}
\label{fracture.area.eps}
\end{figure}

\begin{figure}[tbp]
\includegraphics[width=0.45\textwidth]{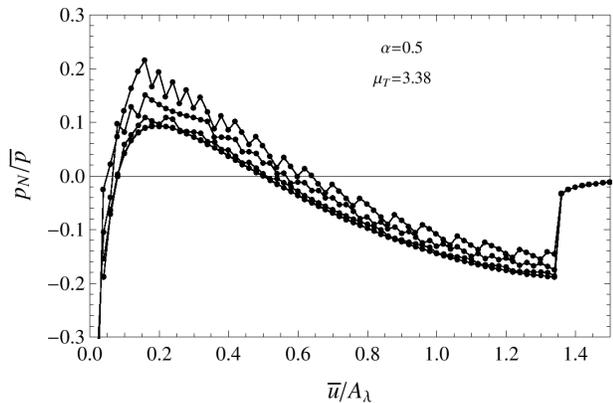}
\caption{{}{}Dimensionless average interfacial separation $\bar{u}/A_{%
\mathrm{\protect\lambda }}$ as a function of the dimensionless applied
pressure $p_{\mathrm{N}}/\bar{p}$, for a Westergaard like contact geometry.
For the same parameters of Fig. \ref{fracture.area.eps}.}
\label{fracture.sep.eps}
\end{figure}

It would be useful to test Eq. (\ref{A1.criterion}) by comparing the Johnson's
solution (Ref. \cite{johnson.sinus}, adhesive sinus contact in the JKR regime) with the
corresponding numerical predictions. In particular, the relation between
nominal contact pressure and contact area reads\cite{johnson.sinus}:%
\begin{equation}
p_{\mathrm{N}}/\bar{p}=\sin ^{2}\phi _{\mathrm{\alpha }}-\alpha \sqrt{\tan
\phi _{\mathrm{\alpha }}},\MSlabel{A1.johnson}
\end{equation}%
where $\phi _{\mathrm{\alpha }}=\pi A/\left( 2A_{0}\right) $, $\alpha =\sqrt{%
2E_{\mathrm{r}}\Delta \gamma /\left( \lambda \bar{p}^{2}\right) }$ and $\bar{%
p}=\pi E_{\mathrm{r}}A_{\mathrm{\lambda }}/\lambda $ (for the adhesionless
interaction, $\bar{p}$ is the nominal squeezing pressure to full contact).
In Fig. (\ref{fracture.area.eps}) we compare Eq. (\ref{A1.johnson}) (red curve)
with numerical results obtained with $\mu _{\mathrm{T,\lambda }}\approx 3.4$
(JKR regime), at different values of detachment parameter $\varepsilon _{%
\mathrm{a}}$. We stress that all the solutions shown in Fig. (\ref%
{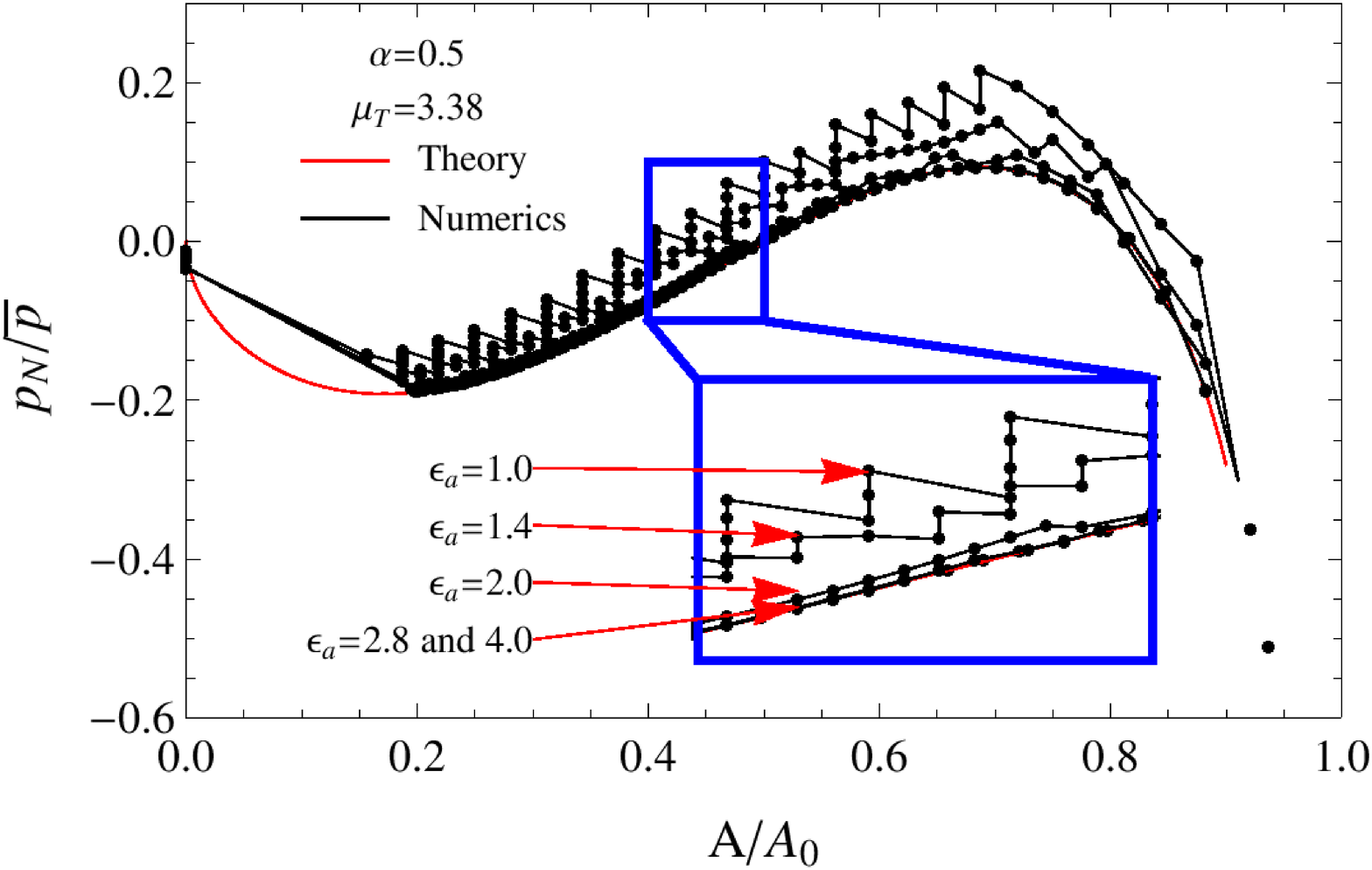}) satisfy the accuracy requirements of Eq. (\ref{A1.err}),
however only at increasing detachment parameter (in particular for $%
\varepsilon _{\mathrm{a}}>2$) the solution rapidly converges to the
analytical one. Similar considerations apply for the pressure-separation
law, shown in Fig. \ref{fracture.sep.eps}.

\vskip 0.3cm \textbf{Acknowledgments} MS acknowledges FZJ for the support
and the kind hospitality received during his visit to the PGI-1, where this
work has been performed. MS also acknowledges COST Action MP1303 for grant
STSM-MP1303-090314-042252 and support from MultiscaleConsulting. 
We thank M.H. M\"user for the useful discussion and comments on the manuscript.


\begin{thebibliography}{99}
\bibitem{Bowden} F.P. Bowden and D. Tabor, \textit{Friction and Lubrication
of Solids} (Wiley, New York, 1956).

\bibitem{Johnson} K.L. Johnson, \textit{Contact Mechanics}, (Cambridge
University Press, Cambridge, 1966).

\bibitem{Isra} J.N. Israelachvili, \textit{Intermolecular and Surface Forces}
(Academic, London (1995)).

\bibitem{BookP} B.N.J. Persson, \textit{Sliding Friction: Physical
Principles and Applications}, 2nd edn. (Springer, Heidelberg, 2000).

\bibitem{R1} B.N.J. Persson, O. Albohr, U. Tartaglino, A.I. Volokitin and E.
Tosatti, J. Phys.: Condens. Matter \textbf{17}, R1 (2005).

\bibitem{P3} B.N.J. Persson, Surface Science Reports \textbf{61}, 201 (2006).

\bibitem{Fuller} K.N.G. Fuller and D. Tabor, Proc. Roy. Soc. London A\textbf{%
345}, 327 (1975).

\bibitem{DMTP} B.V. Derjaguin, Kolloid Z \textbf{69} 155 (1934).

\bibitem{JKRP} K.L. Johnson, K. Kendall and A.D. Roberts, Proc. R. Soc.
Lond. A. \textbf{324}, 301 (1971)

\bibitem{GW}
J.A. Greenwood and J.B.P. Williamson, Proc. Roy. Soc. Lond. A Mat {\bf 295}, 300 (1966).

\bibitem{JG} K.L. Johnson and J.A. Greenwood, J. Colloid Interface Sci. 
\textbf{192}, 326 (1997).

\bibitem{Maugis96} D. Maugis, Journal of Adhesion Science and Technology 
\textbf{10}(2), 161-175 (1996)

\bibitem{Mus3} N. Prodanov,  W.B. Dapp and M.H. M\"user, 
Tribol. Lett. {\bf 53}, 433 (2014).


\bibitem{Mus1} W.B. Dapp, N. Prodanov and M.H. M\"user, \textit{Systematic
analysis of Persson's contact mechanics theory for randomly rough surfaces},
to be published.

\bibitem{Carb1} G. Carbone and F. Bottiglione, J. Mech. Phys. Solids \textbf{%
56}, 2555 (2008).

\bibitem{Comb1} C. Campana, M.H. M\"user and M.O. Robbins, J. Phys.:
Condens. Matter \textbf{20}, 354013 (2008).

\bibitem{Problem} W.B. Dapp, A. L\"ucke, B.N.J. Persson and M.H. M\"user, 
PRL {\bf 108}, 244301 (2012).

\bibitem{mark} L. Pastewka and M.O. Robbins, PNAS \textbf{111}, 3298 (2014)

\bibitem{dini} S. Medina and D. Dini, International Journal of Solids and
Structures (2014) - in press

\bibitem{MP} N. Mulakaluri and B.N.J. Persson, EPL \textbf{96} 66003 (2011).

\bibitem{CarboneS} G. Carbone, M. Scaraggi and U. Tartaglino, Eur. Phys. J E%
\textbf{30}, 65 (2009).

\bibitem{PerssonPRL} A.G. Peressadko, N. Hosoda N and B.N.J. Persson, Phys.
Rev. Lett. \textbf{95}, 124301 (2005).

\bibitem{Krick} B. Lorenz, B.A. Krick, N. Mulakaluri, M. Smolyakova, S.
Dieluweit, W.G. Sawyer and B.N.J. Persson, J. Phys.: Condens. Matter \textbf{%
25}, 225004 (2013).

\bibitem{Krick1} B.A. Krick, J.R. Vail, B.N.J. Persson, W.G. Sawyer,
Tribology Letters \textbf{45}, 185 (2012).

\bibitem{Aut} K. Autumn, MRS Bull. \textbf{32}, 473 (2007).

\bibitem{Heepe} L. Heepe and S. Gorb, Annu. Rev. Mater. Res. \textbf{44},
14.1-14.31 (2014).

\bibitem{PerG1} B.N.J. Persson, J. Chem. Phys. \textbf{118}, 7614 (2003).

\bibitem{PerG2} B.N.J. Persson and S. Gorb, J. Chem. Phys. \textbf{119},
11437 (2003).

\bibitem{cellulose} B.N.J. Persson, C. Ganser, F. Schmied, C. Teichert, R.
Schennach, E. Gilli, U. Hirn, Journ. Phys. C: Condens. Matter \textbf{25},
(2013)

\bibitem{skin} B.N.J. Persson, A. Kovalev, S.N. Gorb, Tribology Letters 
\textbf{50}, 17 (2013).

\bibitem{frog} B.N.J. Persson, Journ. Phys. C: Condens. Matter \textbf{19},
376110 (2007)

\bibitem{coulomb3} B.V. Derjaguin and V. Smilga, J. Appl. Phys. \textbf{38},
4609 (1967).

\bibitem{coulomb4} A.D. Roberts, J. Phys. D: Appl. Phys. \textbf{10}, 1801
(1977).

\bibitem{coulomb1} B.N.J. Persson, M. Scaraggi, A.I. Volokitin, M.K.
Chaudhury, EPL \textbf{103}, 36003 (2013).

\bibitem{coulomb2} K. Br\"ormann, K. Burger, A. Jagota and R. Bennewitz, J.
Adhes. \textbf{88}, 598 (2012).

\bibitem{Dug} D. Maugis, J. Colloid Interface Sci \textbf{150}, 243 (1992).

\bibitem{Bart} E. Barthel, J. Phys. D: Appl. Phys. \textbf{41}, 163001
(2008).

\bibitem{Der} V.M. M\"uller, V.S. Yushenko and B.V. Derjaguin, J. Colloid
Interface Sci. \textbf{77}, 91 (1980).

\bibitem{Gree} J.A. Greenwood, Proc. Roy. Soc. London A\textbf{453}, 1277
(1961).

\bibitem{Mart} M.H. M\"user, Beilstein J. Nanotechnol. \textbf{5}, 419
(2014).

\bibitem{P1} B.N.J. Persson, J. Chem. Phys. \textbf{115}, 3840 (2001).

\bibitem{P4} B.N.J. Persson, Phys. Rev. Lett. \textbf{99}, 125502 (2007)

\bibitem{P2} B.N.J. Persson, Eur. Phys. J E\textbf{8}, 385 (2002).

\bibitem{YP} C. Yang and B.N.J. Persson, J. Phys. Condens. Matter \textbf{20}%
, 215214 (2008).

\bibitem{Carlos} A. Almqqvist, C. Campana, N. Prodanov and B.N.J. Persson,
J. Mech. Phys. Solids \textbf{59}, 2355 (2011).

\bibitem{Layer} B.N.J. Persson, J. Phys.: Condens. Matter \textbf{24},
095008 (2012).

\bibitem{Carbone} G. Carbone, B. Lorenz, B.N.J. Persson and A. Wohlers, Eur.
Phys. J. E\textbf{29}, 275 (2009).

\bibitem{Nyak} P.R. Nayak, J. Lubr. Technol. \textbf{93}, 398 (1971).

\bibitem{comment} In general, in order for elastic instabilities to occur in
contact mechanics and sliding friction the material must be soft enough.
This is most easily understood in the context of a particle in a periodic
potential $U(x)$ and connected to an elastic spring $k$ the free end of
which is moving parallel to the corrugated potential ($x$-direction)
(Tomlinson model of sliding friction). In this case energy will be
dissipated in rapid slip events only if the spring is soft enough (see,
e.g., Ref. \cite{BookP} for a discussion of this point).

\bibitem{layer} B.N.J. Persson, J. Phys.: Condens. Matter \textbf{24},
095008 (2012)

\bibitem{guin} B V. Derjaguin, Kolloid Z. 69155--64 (1934)

\bibitem{congrui} C. Jin, A. Jagota, CY. Hui, J. Phys. D: Appl. Phys.\textbf{%
44}, 405303 (2011)

\bibitem{johnson.sinus} K.L. Johnson, lnt. J. Solids Structures \textbf{32}(3/4), 423430 (1995)
\end{thebibliography}
\end{document}